\documentclass[12pt]{article}
\usepackage{epsf}

\newcommand{\np}[3]{Nucl. Phys. {\bf B#1}, #3 (19#2)}
\newcommand{\pl}[3]{Phys. Lett. {\bf B#1}, #3 (19#2)}
\newcommand{\pr}[3]{Phys. Rev. {\bf D#1}, #3 (19#2)}

\newcommand{\vj}[4]{#1~{\bf #2}, #4 (19#3)}

\newcommand{\be}{\begin{equation}}
\newcommand{\ee}{\end{equation}}
\newcommand{\bea}{\begin{eqnarray}}
\newcommand{\eea}{\end{eqnarray}}
\newcommand{\nn}{\nonumber\\}

\newcommand{\lhs}{{\it l.h.s.\/}} 
\newcommand{\rhs}{{\it r.h.s.\/}} 
\newcommand{\ie}{{\it i.e.\/}}
\newcommand{\eg}{e.g.}

\newcommand{\la}{\langle}
\newcommand{\ra}{\rangle}
\newcommand{\Tr}{\mbox{Tr}}
\newcommand{\Str}{\mbox{Str}}
\newcommand{\N}{{\cal N}}
\renewcommand{\O}{{\cal O}}
\newcommand{\G}{{\cal G}}
\newcommand{\R}{{\cal R}}
\newcommand{\eps}{\varepsilon}
\newcommand{\I}{{\cal I}}

\makeatletter 
\def\secteqno{\@addtoreset{equation}{section}%
\def\theequation{\thesection.\arabic{equation}}}

\begin{document}

\secteqno


\renewcommand{\thefootnote}{\fnsymbol{footnote}}

{\hfill \parbox{4cm}{ 
        UCLA/2000/TEP/12 \\ 
        MIT-CTP-2964 \\ 
        hep-th/0003218 \\
        March 2000                  
}}

\bigskip\bigskip

\begin{center} \Large \bf 
Near-extremal correlators and \\
vanishing supergravity couplings  in AdS/CFT 
\end{center}

\bigskip\bigskip

\centerline{ Eric D'Hoker$^{a}$, Johanna Erdmenger$^b$, 
Daniel Z. Freedman$^{b,c}$ and 
Manuel P\'{e}rez-Victoria$^{b}$\footnote[1]{\tt 
dhoker@physics.ucla.edu, jke@mitlns.mit.edu, 
dzf@math.mit.edu, manolo@pierre.mit.edu} }

\bigskip
\bigskip
\centerline{$^a$ \it Department of Physics}
\centerline{ \it University of California, Los Angeles, CA 90095}
\bigskip
\centerline{$^b$ \it Center for Theoretical Physics}
\centerline{ \it Massachusetts Institute of Technology}
\centerline{ \it Cambridge, {\rm MA}  02139}
\bigskip
\centerline{$^c$ \it Department of Mathematics}
\centerline{ \it Massachusetts Institute of Technology}
\centerline{\it Cambridge, {\rm MA} 02139}
\bigskip\bigskip

\bigskip

\renewcommand{\thefootnote}{\arabic{footnote}}

\centerline{\bf Abstract} 
\medskip 

We study near-extremal $n$-point correlation functions of chiral primary
operators, in which the maximal scale dimension $k$ is
related to the others by $k=\sum_i k_i-m$ with $m\leq
n-3$. Through order $g^2$ in field theory, we show that these
correlators are simple sums of terms each of which factors into
products of lower-point correlators. Terms which contain only factors
of two- and three-point functions are not renormalized, but other
terms have non-vanishing order $g^2$ corrections. \\ 
\indent
We then show that the contributing AdS exchange diagrams neatly match
this factored structure. In particular, for $n=4,5$ precise agreement in
form and  
coefficient is established between supergravity and the
non-renormalized factored terms from field theory. On the other hand,
contact diagrams in supergravity would produce a non-factored
structure. This leads us to conjecture that the corresponding bulk
couplings vanish, so as to achieve full agreement between the
structure of these correlators in supergravity and weak-coupling field
theory. 

\newpage


\section{Introduction}

The AdS/CFT correspondence \cite{Maldacena,Gubser,Witten} 
has enabled the exact calculation 
of many correlation functions in a strong coupling limit of $\N$=4 
$SU(N)$ 
supersymmetric Yang-Mills theory  and to the initially suprising fact 
that many correlators appear to be non-renormalized-- strong coupling results
agree with free field limits. Although it is debatable whether a rigorous
proof has been achieved, there is ample evidence from an interplay of
arguments from AdS supergravity, order $g^2$ and $g^4$ calculations and
formal non-perturbative considerations in the field theory that 
non-renormalization holds for all values of $g$ and $N$ and for
general gauge groups.

This line of investigation is continued in the present paper in which we
present new results on the structure of near-extremal $n$-point
correlators. We consider the chiral primary operators $\O_k =  \Tr X^k$  and
study order $m$ sub-extremal $n$-point functions 
$\la \O_k\O_{k_1}\cdots\O_{k_{n-1}} \ra$ with $k=k_1+\cdots+k_{n-1}-2m$
and $0 \leq m \leq n-3$. We shall call such correlators $E_n^m$ functions.
As we will review in more detail below, previous studies strongly suggest that
(extremal) $E_n^0$ and (next-to-extremal) $E_n^1$ functions for $n\geq
3$ are not renormalized. 
On the field theory side these correlators are characterized
by the factorization of their free-field graphs into products
of two- and three-point structures. Order $g^2$ radiative corrections
and Yang-Mills 
instanton corrections vanish \cite{Bianchi,EP}. For $n\geq 4$ 
the contributing 
exchange diagrams from Type
IIB supergravity on $AdS_5\times S^5$ reproduce the factored
space-time form, but contact diagrams involving quartic or higher
order vertices do not. Thus  
supergravity $E_n^0$ and $E_n^1$ couplings should vanish, and this has been
verified \cite{Frolovnew} for $n=4$.

For near-extremality, $m\geq 2$, the situation is more complex as we now
exemplify for 
the case $m=2$. Here the $E_4^2$ correlator 
$\la\O_2\O_2\O_2\O_2\ra$ is known to be renormalized to
order $g^2$ in field theory \cite{Rey, Westf} (for a recent calculation 
to three loops see \cite{ESS}) and in the large $N$ strong coupling
limit in supergravity \cite{Frolov4222,Petkou,Chalmers}. 
Indeed $E_4^2$ correlators have no
special factored free-field limits, which suggests that one must examine
$E_5^2$ functions to find non-renormalization. It is easy to see that the
free-field graphs for the $E_5^2$ function 
$\la\O_4\O_2\O_2\O_2\O_2\ra$ are of two distinct types;
see Fig.~\ref{fig_free42222}. The first is a product of two three-point
structures while the second 
is a product of a two-point and four-point structure. We show that the
order $g^2$ 
radiative corrections to the first type vanish, but corrections to the
second type do not vanish as might be expected because there is a four-point
sub-structure present. Therefore the party is over as far as complete 
non-renormalization is concerned, but the factored structure is
preserved by radiative corrections and this suggests that
we look at the situation in supergravity.

Indeed it is very striking that the structure found at weak coupling is
exactly mirrored at strong coupling in supergravity. In particular one of the
two contributing double-exchange diagrams in supergravity reproduces the
first factored structure of free field theory {\bf both in space-time form and 
with exactly the same numerical strength}. The second double exchange and
both single exchange diagrams reproduce the second factored structure from
field theory, but contain a non-trivial four-point sub-structure which cannot 
be directly compared with weak coupling. Full agreement in structure between
supergravity and weak field theory requires that the supergravity five-point
coupling corresponding to the
$s_{4}s_2s_2s_2s_{2}$ vertex vanishes. 
And indeed it must vanish for consistent
decoupling of the multiplet containing the bulk field
$s_4$ dual to $\O_4$ from the graviton multiplet. This is required by 
consistent Kaluza-Klein truncation of Type IIB supergravity on $AdS_5
\times S^5$, which means that the solutions to
the equations of motion of $\N$=8 gauged supergravity on $AdS_5$ are
exact solutions of the 
complete Type IIB supergravity theory on $AdS_5 \times S^5$
\cite{grw,ppvn}. This property forbids terms in the full
action which both contain the graviton multiplet and are linear in a
field of a higher Kaluza-Klein multiplet.

We show that the
example of $\la\O_4\O_2\O_2\O_2\O_2\ra$ generalizes
to all $E_5^2$ functions (provided an extra assumption about
descendent couplings holds). The factored structure of weak-coupling field
theory agrees with supergravity calculations if all $E_5^2$ bulk couplings
vanish, which is our prediction.
There is a further extension of these results to all $E_n^m$ functions with
$m\leq n-3$: Supergravity reproduces  the factored space-time form
of the field-theoretical calculation if bulk $E_n^m$ couplings
vanish. This leads us to conjecture that these couplings do indeed
vanish. 

We shall now give a detailed summary of the non-renormalization properties
of $\N$=4 SYM correlators which have been established through study
of the AdS/CFT correspondence.

{\em 1. Two- and three-point functions of single-trace chiral
operators.\/} The 
first result of this type came in \cite{Seiberg} where a calculation
of cubic couplings $g_{k_1k_2k_3}$ of the bulk fields $s_k$ dual to the
operators $\O_k$ was combined with results \cite{Freedman3pt} for AdS
three-point integrals. Operators were normalized to have unit
two-point functions, and 
it was observed that supergravity results for all three-point functions
$\la\O_{k_1}\O_{k_2}\O_{k_3}\ra$ agreed with free field theory. This
was soon followed by an explicit study of order $g^2$ radiative corrections
in field theory which were shown to vanish for both two-point functions
$\la\O_{k}\O_{k}\ra$ and three-point functions of non-normalized operators
\cite{FreedmanCFT}. A superspace calculation may be found in
\cite{Rey2}. Recently it was shown \cite{Zanon} that order $g^4$
contributions to $\la\O_3\O_3\ra$ vanish. 

There have been many attempts to prove the
non-renormalization of the three-point functions non-perturbatively in field
theory. All approaches require unproved technical assumptions. Most convincing
are the arguments using $\N$=2 analytic superspace \cite{West1,West2} that 
there are
no possible superspace forms which could contribute to the derivative of a
three-point function \cite{Intriligator} with respect to the gauge coupling.
(See also \cite{PS}).

{\em 2. Two- and three-point functions of other chiral operators.\/} Related
results have emerged from studies \cite{Ferrara} of short  
representations \cite{Dobrov} of the conformal superalgebra $SU(2,2|4)$ of
the $\N$=4 SYM theory. The chiral primaries $\O_k$ are 
1/2-BPS operators which transform in the
$[0,k,0]$ representation of the $R$-symmetry group $SU(4)$. There are other
1/2-BPS multitrace operators in the same representations, such as the
projection of $:{\rm Tr}X^{k_1}{\rm Tr}X^{k_2}:$ 
in the $[0,k_1+k_2,0]$ representation,
and all of them have protected dimension $\Delta_k=k$. It is therefore of
interest to check the non-renormalization properties of these operators,
and it was shown in \cite{Skiba} that order $g^2$ radiative corrections to
their two- and three-point functions vanish for all gauge groups. The
same question can be asked of the 1/4-BPS 
operators of dimension $p+2q$ in the $SU(4)$ representation $[q,p,q]$
and the 1/8-BPS operators of dimension $p+2q+3r$ in the representation
$[q,p,q+2r]$. This is work in progress. See also~\cite{Bianchi2,Bianchi3}.

{\em 3. Extremal and next-to-extremal functions.\/} The study of
$E_n^{0,1}$ functions of chiral primaries for 
$n\geq4$ emerged from a peculiarity of  
$E_3^0$ functions noted in \cite{Freedmanextremal}. It was shown that
the factored 
form of free field theory was reproduced in supergravity as the product of
a zero bulk coupling constant with an infinite AdS integral. The product
was defined by analytic continuation in the dimensions $\Delta_k$ of the
operators, and this procedure was justified in a related example by careful
consideration of boundary interactions. It was then noticed 
\cite{Freedmanextremal}
that $E_n^0$ functions had factored space-time forms for all $n\geq4$. 
All contributing supergravity diagrams, 
each defined by analytic continuation, give the same factored form. The
coefficient of this form could then be shown to vanish by OPE arguments.
After order $g^2$ perturbative test in \cite{Bianchi}, non-perturbative
$\N=2$ and $\N=4$ superspace arguments were given to support this result and
to suggest that $E_n^1$ next-to-extremal correlators are not renormalized
\cite{West}. This was then shown for any $n$ through order $g^2$
\cite{EP} and in AdS supergravity~\cite{EP,Frolovnew}.

{\em 4. Near-extremal functions.\/} This brings us to the situation of
$E_n^m$ functions for $2\leq m \leq n-3$
which we have summarized at the  beginning of this introduction with
supporting arguments to be given below.

The paper is organized as follows. In Section 2 we present our results by
considering the simplest possible case, which is the correlator
$\la \O_4 \O_2 \O_2 \O_2 \O_2\ra$. We then move on to general 
$E^m_n$ functions, which we discuss from 
the field theory point of view in Section 3 and from the supergravity side
in Section 4, paying particular attention to  supergravity couplings
in Section 5. There is a short conclusion and an Appendix which contains
a detailed consideration of the AdS calculations essential to our argument.


\section{The correlator $\la \O_4 \O_2 \O_2 \O_2 \O_2
\ra$}

In order to illustrate the structure of $E^2_n$
correlation functions in the simplest possible situation, we study in
detail the
correlator $\la \O_4(x) \O_2(x_1) \O_2(x_2) \O_2(x_3)
\O_2(x_4) \ra$.  We first calculate the
order $g^2$ corrections in SYM and then the corresponding diagrams in
AdS supergravity. Although non-trivial radiative
corrections appear at order $g^2$, their factored form is
compatible with supergravity provided the bulk $s_4s_2s_2s_2s_2$ coupling
vanishes, as required by consistent Kaluza-Klein truncation.

\subsection{$\la \O_4 \O_2 \O_2 \O_2 \O_2 \ra$ in SYM}
\label{subsec_field42222}

We shall use the methods of~\cite{EP} (where the interested reader
can find more details). We normalize the operators as
in~\cite{Seiberg}:
\be
\O_k(x) = \frac{(2\pi)^k}{N^{k/2} \sqrt{k}} \Tr X^k(x) \, . \label{opnorm}
\ee
With this normalization the two point function is given by
\be
\la \O_k(x) \O_k(y) \ra = \frac{1}{(x-y)^{2k}}
\ee

\begin{figure}[ht]
\begin{center}
\epsfxsize=9cm
\epsfbox{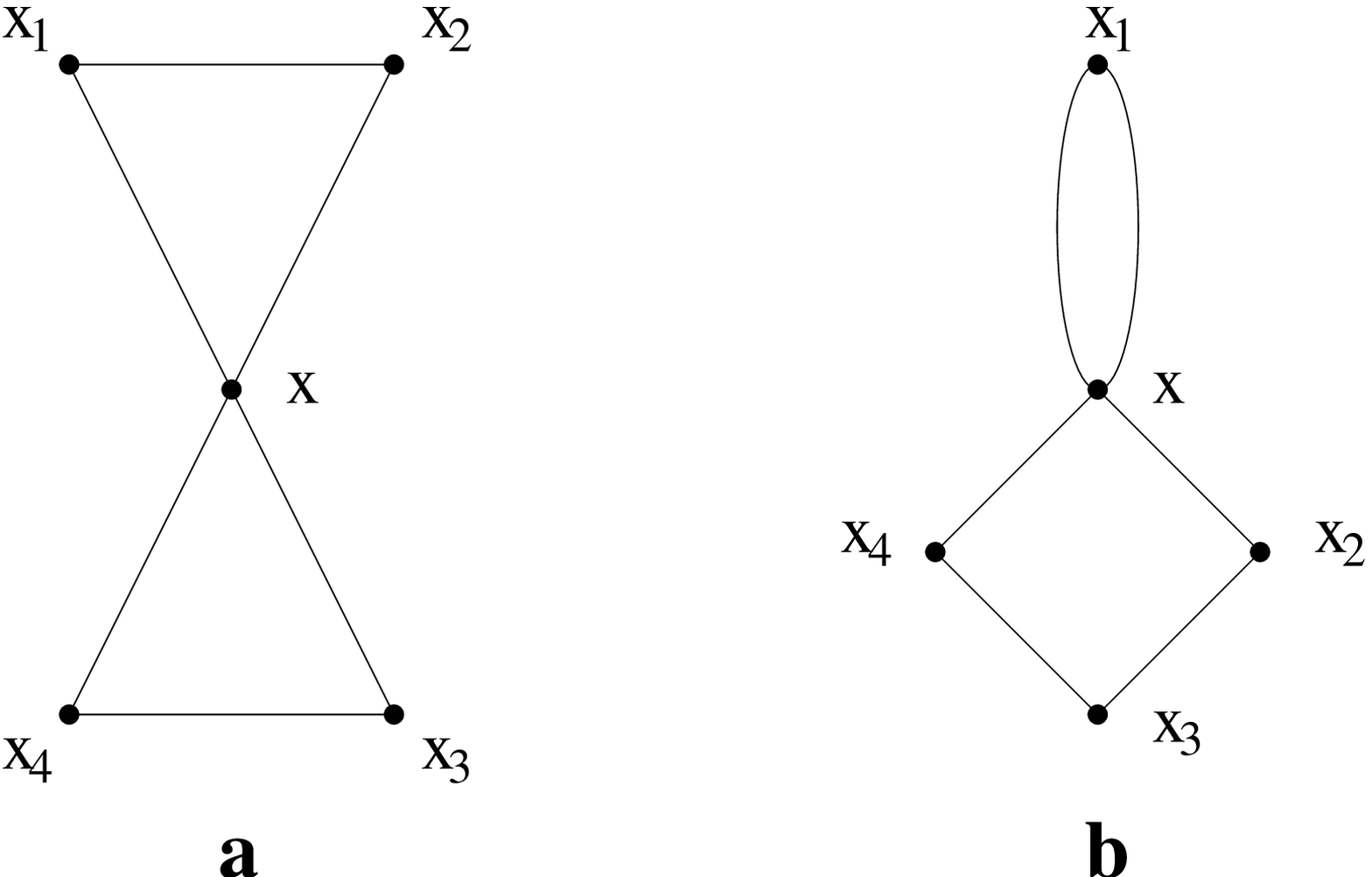}
\end{center}
\caption{Feynman graphs contributing to the correlator $\la \O_4
\O_2 \O_2 \O_2 \O_2 \ra$ at the free-field level. \label{fig_free42222}}
\end{figure}
At the free-field level there are two connected graphs, up to
permutations of the operators $\O_2$ (Fig.~\ref{fig_free42222}). 
Graphs $a$ and $b$  give the contributions
\bea 
\lefteqn{
\la \O_4(x) \O_{2}(x_1) \O_{2}(x_2) \O_{2}(x_3) \O_{2}(x_4) \ra_a} &&
\nn 
&& = {\cal C}^{(a)} Q \frac{1}{(x-x_1)^2} \frac{1}{(x_1-x_2)^2}
\frac{1}{(x_2-x)^2} \frac{1}{(x-x_3)^2} \frac{1}{(x_3-x_4)^2}
\frac{1}{(x_4-x)^2}  ,\label{1b} \\ 
\lefteqn{
\la \O_4(x) \O_{2}(x_1) \O_{2}(x_2) \O_{2}(x_3) \O_{2}(x_4) \ra_b} && \nn
&& = 
{\cal C}^{(b)} Q \frac{1}{(x-x_1)^4} \frac{1}{(x-x_2)^2}
\frac{1}{(x_2-x_3)^2} \frac{1}{(x_3-x_4)^2} \frac{1}{(x_4-x)^2} , 
\eea
respectively,
where ${\cal C}^{(a,b)}$ are tensors in flavour space, and
\bea
Q & = & \Str(T^{a_1}\cdots T^{a_4})
\Str(T^{a_1}T^{a_2}) \Str(T^{a_3}T^b) \Str(T^b T^c) \Str(T^cT^{a_4})
\nn
& = & \Str(T^{a_1}\cdots T^{a_4})
\Str(T^{a_1}T^{b}) \Str(T^bT^{a_2}) \Str(T^{a_3} T^c) \Str(T^c
T^{a_4}) \nn
& = & \frac{N^2-1}{2^6}. 
\eea
The symmetric trace is defined as
\be
\Str(T^{a_1} \cdots T^{a_k}) = \sum_{\mbox{\scriptsize perms}\,\sigma} 
\frac{1}{k!} \Tr(T^{a_{\sigma(1)}} \cdots T^{a_{\sigma(k)}}).
\ee
In the following we suppress explicit flavour tensors. Our order $g^2$ 
calculations are valid for all $N$, but for later comparison 
with supergravity, we write the large $N$ limit of all non-vanishing contributions. Thus we have, for example, 
for
the contribution of the graph $a$ in Fig.~\ref{fig_free42222},
\bea \lefteqn{
\la \O_4(x) \O_{2}(x_1) \O_{2}(x_2) \O_{2}(x_3) \O_{2}(x_4) \ra_a} && \nn
&=& \frac{4}{N}  
\, \la \O_2(x) \O_2(x_1)\O_2(x_2) \ra\, \la \O_2(x)\O_2(x_3)
\O_2(x_4) \ra, \label{tb} 
\eea 
where
\be
\la \O_2(x) \O_2(x_1)\O_2(x_2) \ra
 = \frac{2\sqrt{2}}{N} \frac{1}
{ (x-x_1)^2 (x_1-x_2)^2 (x_2-x)^2} \, . \label{t8}
\ee
The coefficients in (\ref{tb}) and (\ref{t8}) incorporate both Wick
combinatoric factors and the normalization factors of (\ref{opnorm}).

\begin{figure}[ht]
\begin{center}
\epsfxsize=12cm
\epsfbox{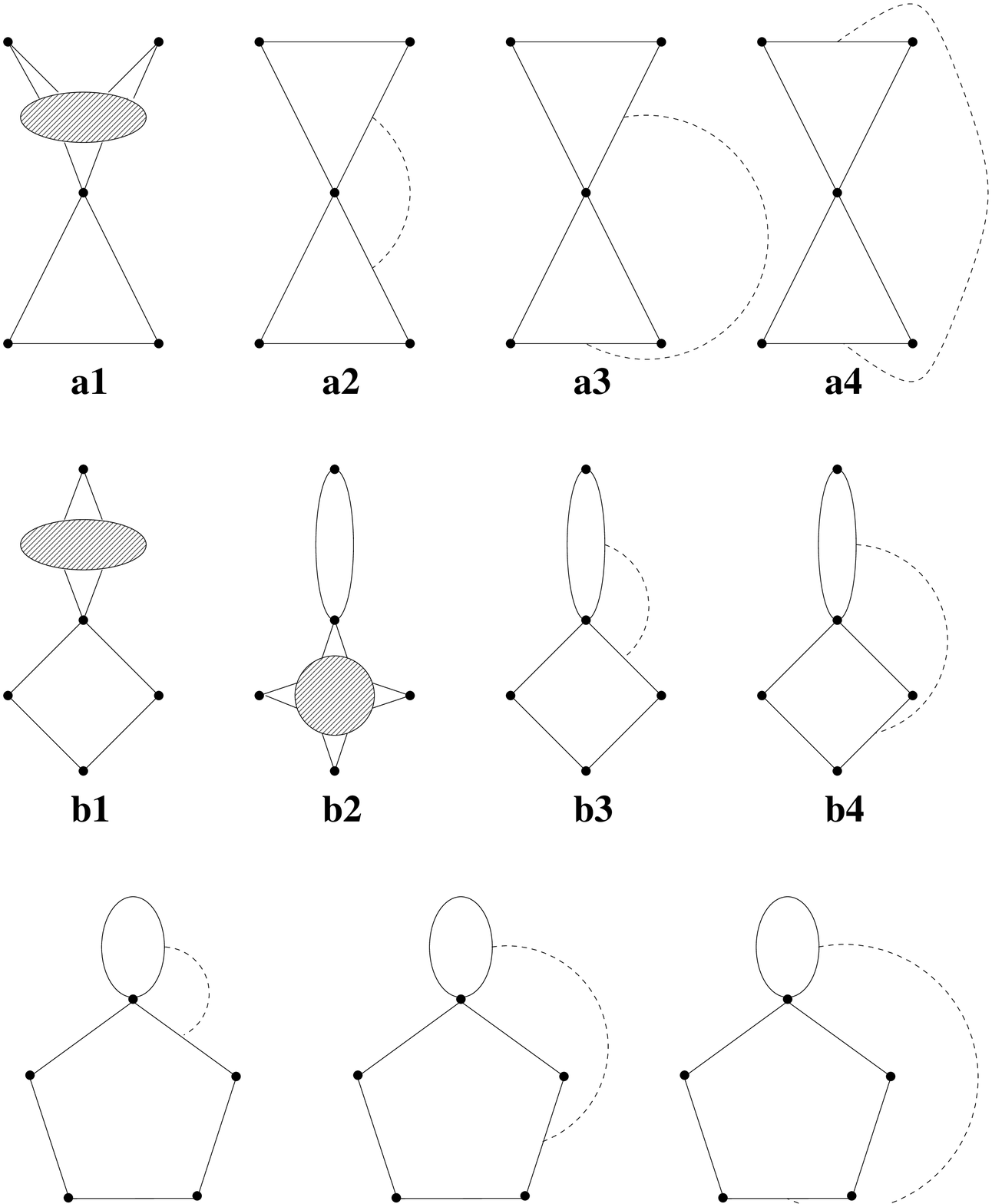}
\end{center}
\caption{Graphs contributing to the correlator $\la \O_4
\O_2 \O_2 \O_2 \O_2 \ra$ at order $g^2$.  \label{fig_g42222}}
\end{figure}
Let us now consider
the connected Feynman graphs contributing at order $g^2$, which are
depicted in Fig.~\ref{fig_g42222}. The dashed lines denote the
combination of gauge boson exchanges and quartic scalar interactions
(see \cite{EP} for details).
Graphs $a1$, $b1$ vanish due to
the well-known non-renormalization theorems for the functions $\la \O_2
\O_2 \ra$  and $\la \O_2 \O_2 \O_2 \ra$. These theorems hold independently of 
colour contractions, see \cite{Skiba} and the further explanations 
below Eq. (13) in \cite{EP}. 
All the remaining graphs but $b2$ vanish due to the
symmetry properties of colour indices. Let us show this explicitly for
$a3$:
\bea
a3 &\sim & \Str(T^{a_1}\cdots T^{a_4}) \Str(T^bT^c)
\Str(T^c T^{a_4}) \Str(T^dT^{a_2}) \Str(T^eT^{a_3}) \nn
&& \hspace{.5cm} \cdot \; \left(A^{(a3)} f^{a_1bp}f^{dep} + B^{(a3)}
f^{a_1dp}f^{bep} 
+ C^{(a3)} 
f^{a_1ep} f^{bdp} \right) \nn
 & = & \Str(T^{a_1}\cdots T^{a_4}) \left( A^{(a3)}
f^{a_1a_4p}f^{a_2a_3p} + B^{(a3)} f^{a_1a_2p}f^{a_4a_3p} + C^{(a3)}
f^{a_1a_3p} f^{a_4a_2p} \right),  
\eea
which vanishes as each term is a contraction of a symmetric with an
antisymmetric tensor. Graphs $a2$, $a4$, $b3$, $b4$, $c1$, $c2$ and
$c3$ can be shown to vanish in the same way. Therefore, the only
possible contribution is that of graph b2, which factors into a
two-point and a four-point function. For large $N$, the contribution
of $b2$ is given by
\bea \lefteqn{
\la \O_4(x) \O_{2}(x_1) \O_{2}(x_2) \O_{2}(x_3) \O_{2}(x_4) \ra_{b2}}
&& \nn
&=& \frac{4}{N} \, 
{\la \O_2(x)\O_2(x_1) \ra}  {\la \O_2(x)\O_2(x_2)
\O_2(x_3)\O_2(x_4) \ra}^{(1)}, \label{faca2}
\eea
where the index $(1)$ indicates order $g^2$. We do not indicate the
order in two- and three-point functions as they are not renormalized.
The second factor was calculated and shown to contain logarithms in
\cite{Rey,Westf,ESS}.  
Therefore the full five-point correlator is
renormalized at order $g^2$. Nevertheless, the contribution does factor.
Hence to order $g^2$, the complete $E_5^2$ correlator
is given by the sum of the two factored contributions (\ref{tb}) 
and (\ref{faca2}),
\bea 
\lefteqn{\la \O_4(x) \O_2(x_1) \O_2(x_2) \O_2(x_3) \O_2(x_4) \ra} && \nn 
 &=& \frac{4}{N}   
\, \la \O_2(x) \O_2(x_1)\O_2(x_2) \ra\, \la \O_2(x)\O_2(x_3)
\O_2(x_4) \ra \nn 
& & \mbox{} + \frac{4}{N}
{\la \O_2(x)\O_2(x_1) \ra}  {\la \O_2(x)\O_2(x_2)
\O_2(x_3)\O_2(x_4) \ra} \,
+ \mbox{perms.}  \, .
\label{fac42222}
\eea
The two-point and three-point structures are not
renormalized. This factored
space-time structure is a consequence of
the properties of $\N=4$ SYM. Conformal invariance alone
permits a more general structure.

\subsection{$\la \O_4 \O_2 \O_2 \O_2 \O_2
\ra$ in AdS}
\label{subsec_AdS42222}

\begin{figure}[ht]
\begin{center}
\epsfxsize=15cm
\epsfbox{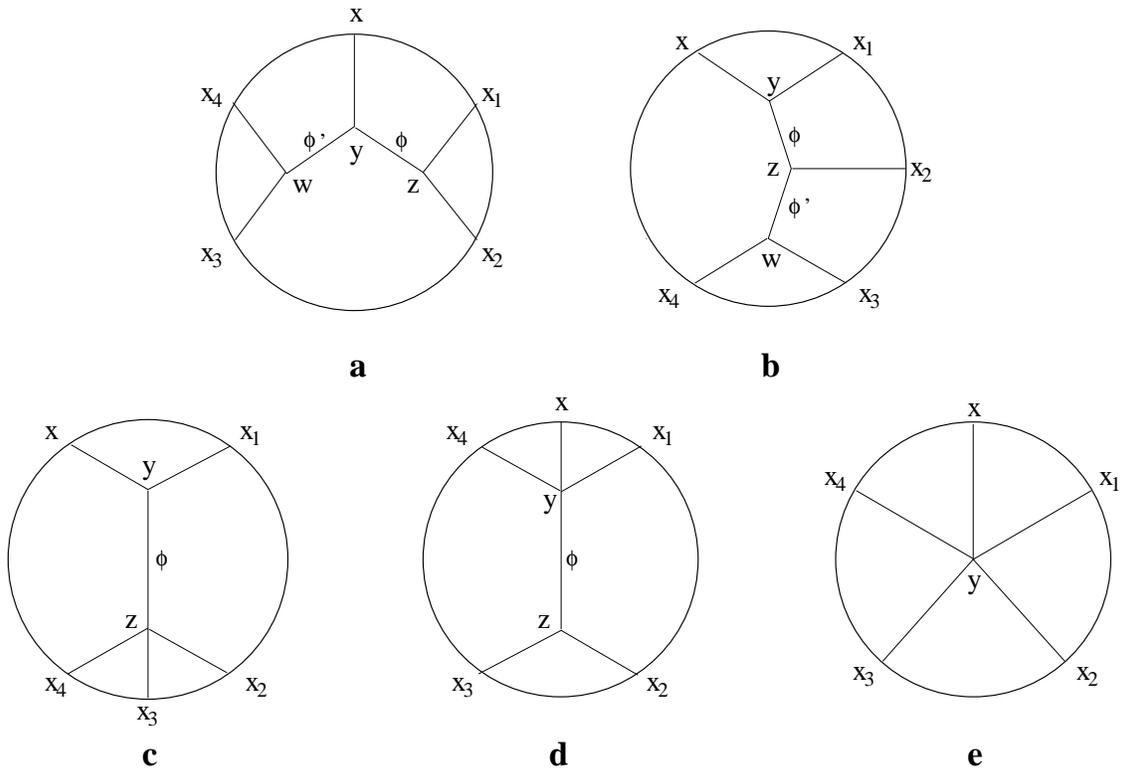}
\end{center}
\caption{Witten diagrams contributing to $E_5^2$ correlators in the AdS/CFT
correspondence. \label{fig_AdS42222}} 
\end{figure}

According to the Maldacena conjecture, the same function can be
calculated (at strong coupling and large $N$) using classical Type IIB
supergravity on $AdS_5\times S^5$. 
The corresponding connected Witten
diagrams are shown in Fig.~\ref{fig_AdS42222}, up to permutations.
We refer to the bulk fields dual to the operator $\O_k$ as $s_k$.
There are double-exchange ($a$ and   
$b$), single-exchange ($c$ and $d$) and contact diagrams ($e$). 
The cubic and quartic couplings we need were calculated in
\cite{Seiberg, Frolov3, Lee} and \cite{Frolov4}, respectively. The
main property we need for our purposes is that the couplings of the
vertices $s_2s_2s_4$, $s_2s_2s_2s_6$ and $s_2s_2s_2s_4$ vanish 
(although allowed by $SU(4)$ symmetry). By ``vanishing coupling'' we
mean that the combination of derivative and non-derivative terms in
the vertex give a vanishing net coupling in the bulk. 
It is important to note that even if the coupling
vanishes in the bulk, there can be a contribution from a surface term if
the space-time integral is divergent~\cite{Freedmanextremal}. 
In \cite{Freedmanextremal} it
was shown (for a particular case) that this boundary
contribution is also obtained when couplings and integrals
are regularized by analytic continuation. Moreover, 
analytic continuation has successfully produced results which agree
with field theory~\cite{Seiberg, Freedmanextremal, EP}. We use
this method here as well and  
regularize the divergent integrals 
by analytic continuation in the highest conformal dimension,
$k=\Delta=4 \rightarrow 4-\eps$, which also implies 
that the vanishing couplings give rise to factors of 
$\eps$.\footnote{An alternative procedure
would be to analyze the implications of field redefinitions removing  
the surface terms, in the spirit
of~\cite{Frolovnew}.}
The convention of normalized two-point functions \cite{Seiberg}
requires that AdS integrals must be divided by a product of factors
\bea 
\N_{k} & = & 
\frac{4N}{\pi^2} \frac{1}{2^{k/2}} \frac{\sqrt{k}(k-1)(k-2)}
{(k+1)}, ~ k>2 \, , \nn
\N_{2} & = &  \frac{4N}{\pi^2} \frac{1}{\sqrt{2}\cdot 3} \, ,
\label{AdSnorm} 
\eea
one for each external $s_k$ line\footnote{The normalization for $k=2$
is in agreement with the discussion in~\cite{KW}.}.
For the couplings $\G$ 
we also use the expressions given in \cite{Seiberg}, and the Poisson kernels
$K_k(x,z)$ are given in the Appendix.

Let us first consider the exchange of primary fields.
We begin with diagram $a$.  The exchanged fields $\phi$, $\phi'$
have to be in the representation with Dynkin labels
$[0,\delta,0]$, $[0,\delta',0]$, respectively, with $\delta$, $\delta'$
restricted to $2$ and $4$ in the present example. 
The contribution of the exchange diagram for generic $\delta$, $\delta'$ 
is given by
\bea
\lefteqn{
\la \O_4(x) \O_2(x_1) \O_2(x_2) \O_2(x_3) \O_2(x_4) \ra_a } && \nn 
&& = \frac{{\cal G}(4,\delta,\delta^\prime){\cal
G}(\delta,2,2){\cal
G}(\delta^\prime,2,2)}{\N_{4}\N_{2}{}^4} \;
\int \!\!\!\int \!\!\!\int  \, \frac{ d^5y}{y_0^5}\frac{ d^5z}{z_0^5}
\frac{d^5w}{w_0^5} \; K_4(x,y) G_\delta (y,z) 
\nn 
&& \hspace{1cm} \cdot \;
K_{2}(x_1,z) K_{2} (x_2,z) G_{\delta'}(y,w) K_2(x_3,w)
K_{2} (x_4,w) \, . \label{5b}
\eea
\begin{figure}[ht]
\begin{center}
\epsfxsize=11cm
\epsfbox{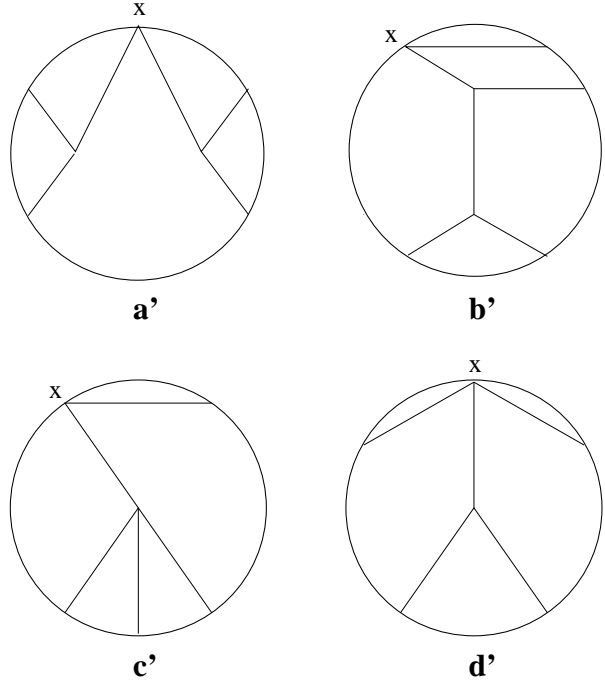}
\end{center}
\caption{Non-vanishing contributions to $E_5^2$ functions in AdS
supergravity. Diagram 
$d'$ does not appear for $\la \O_4
\O_2 \O_2 \O_2 \O_2 \ra$. \label{fig_red42222}}
\end{figure}
If $\delta=\delta'=2$, the vertex
$y$ is extremal, such that the corresponding coupling vanishes \cite{Seiberg},
and it is adjacent to the boundary operator with 
the highest conformal dimension. It is crucial to note here and for the 
subsequent discussion that in this case the integral over $y$ is divergent,
with the dominant contribution arising when $y \sim x$. 
As shown in detail in the Appendix, using analytic continuation $\Delta
\rightarrow \Delta- \varepsilon$ for the highest conformal dimension, both
for the extremal coupling and to evaluate the $y$ integral, 
we find an unambiguous finite expression for
$\eps\rightarrow 0$.

The remaining integrals 
over $z$ and $w$ are finite and, as represented in
diagram $a'$ of Fig.~\ref{fig_red42222}, factor into two three point
functions 
\bea
\la \O_2(x)\O_2(x_1)\O_2(x_2) \ra = \frac{\G(2,2,2)}{ \N_{2}{}^3 }    \, 
\int \! \frac{d^5z}{z_0^5} \; K_2(x,z)
K_2(x_1,z) K_2(x_2,z) \, ,
\eea 
which may be evaluated using \cite{Freedman3pt} with the result 
\bea
\la \O_2(x)\O_2(x_1)\O_2(x_2) \ra
 = \frac{2 \sqrt{2}}{N} \frac{1}
{ (x-x_1)^2 (x_1-x_2)^2 (x_2-x)^2} \, .
\eea
Putting all the factors together and inserting the explicit values for the
supergravity couplings, we obtain for the five-point function
contribution (\ref{5b}) 
\bea \lefteqn{
\la \O_4(x) \O_{2}(x_1) \O_{2}(x_2) \O_{2}(x_3) \O_{2}(x_4) \ra_a}
&& \nn&=& \frac{4}{N}  
\, \la \O_2(x) \O_2(x_1)\O_2(x_2) \ra\, \la \O_2(x)\O_2(x_3)
\O_2(x_4) \ra \, ,
\label{tbAdS}
\eea
which is in {\bf exact agreement in form and value} 
with the free-field contribution (\ref{tb}).
Later  we show that this is valid for general conformal dimensions as well,
not only in the simple case considered here as an introduction.
This is both a non-trivial test of the AdS/CFT correspondence and of
the method of analytic continuation.
Of course
this explicit comparison is possible since the three point functions
involved are not renormalized, and there is exact agreement  
of the functional form between weak and strong coupling. As far as the
factors 
are concerned, we see that the field theory large $N$ 
Wick factors agree with the
factors obtained from the supergravity couplings, the normalization of the
propagators and the integral evaluated using analytic continuation.

If $\delta+\delta^\prime>4$ in (\ref{5b}), at least one of the remaining
vertices is extremal and its coupling vanishes. 
The integral, however,  is finite and hence the diagram
vanishes.

Let us now consider diagram $b$, where the exchanged field $\phi$
has to be in the representation
$[0,\delta,0]$ with $\delta=2,4,6$ 
and conformal dimension $\delta$. If $\delta=2$, the cubic vertex at
$y$ is again extremal and the corresponding coupling vanishes.
Again, there is a zero-times-infinity situation, 
such that a finite contribution 
emerges from the region $y\sim x$:
\bea
\lefteqn{\la \O_4(x) \O_{2}(x_1) \O_{2}(x_2) \O_{2}(x_3) 
O_{2}(x_4) \ra_b} \hspace{1cm} && \nn
 &=& \frac{1}{\N_{4}\N_{2}^4} \,
\sum_{\delta^\prime=2,4} {\cal G}(4,2,2){\cal
G}(\delta^\prime,2,2){\cal G}(\delta^\prime,2,2) \;
\int\!\!\int\!\!\int \!
\frac{d^5y}{y_0^5}\frac{d^5z}{z_0^5}\frac{d^5w}{w_0^5} 
K_4(x,y) K_2(x_1,y) \nn
&& \hspace{0.5cm} \cdot \; G_2(y,z) K_2(x_2,z)
G_{\delta^\prime}(z,w)K_2(x_3,w)K_2(x_4,w) \nn 
& = & \frac{2 \sqrt{2}}{N} \, \la \O_2(x) \O_2(x_1) \ra \,
\frac{1}{\N_{2}^3}
\sum_{\delta^\prime=2,4}{\cal G}(\delta^\prime,2,2)
{\cal G}(\delta^\prime,2,2)  \nopagebreak \nn
&& \hspace{0.5cm} \cdot \; \int\!\!\int  \!
\frac{d^5z}{z_0^5}\frac{d^5w}{w_0^5} 
K_2(x,z) K_2(x_2,z) G_{\delta^\prime}(z,w)K_2(x_3,w)K_2(x_4,w) \nn
& = & \frac{4}{N} \la \O_2(x) \O_2(x_1) \ra
\la \O_2(x) \O_2(x_2) \O_2(x_3) \O_2(x_4) \ra_b
. \label{red42222a}
\eea
Again there is agreement between the factors in (\ref{red42222a})
and in (\ref{faca2}). However since the four-point functions are
renormalized, 
it is not possible to compare them directly: In (\ref{faca2}) we have
an order $g^2$ contribution at weak coupling, whereas in 
(\ref{red42222a}) we have an exchange contribution at strong coupling.
Diagram $b'$ of
Fig.~\ref{fig_red42222} represents the space-time structure of
Eq.~(\ref{red42222a}). The integrals
on $z$ and $w$ are finite. 

If $\delta=4$ (such that the vertex at $y$ is next-to-extremal), the
field $\phi^\prime$ is in the representation $[0,\delta^\prime,0]$ with
$\delta^\prime=2$ or $\delta^\prime=4$. Then either the 
vertex at $z$ or the vertex at $w$ is extremal and hence has a
vanishing coupling. If
$\delta=6$, the only possibility is $\delta^\prime=4$ and the three
vertices are extremal. On the other hand, as shown in the Appendix, the
integrals are finite in all these cases. Therefore the corresponding
diagrams vanish. 

The four-point factor in (\ref{red42222a})
corresponds to an exchange diagram. In diagram $c$ we can have
$\delta=2,4,6$. If $\delta=2$ the vertex $y$ is extremal and the
diagram factors:  
\bea
\lefteqn{\la \O_4(x) \O_{2}(x_1) \O_{2}(x_2) \O_{2}(x_3) 
\O_{2}(x_4) \ra_c} && \nn
& = & \frac{2 \sqrt{2}}{N} \la \O_2(x) \O_{2}(x_1) \ra \,
{\cal G}(2,2,2,2) \int \! \frac{d^5z}{z_0^5} K_2(x,z)
K_2(x_2,z) K_2(x_3,z) K_2(x_4,z) \nn
&=&  \frac{4}{N} \la \O_2(x) \O_{2}(x_1) \ra \, 
  {\la \O_2(x)\O_2(x_2)
\O_2(x_3)\O_2(x_4) \ra}_c \, ,
\eea
where the four-point function is the contact contribution to the
$E_4^2$ function. 
The factored structure is shown in
diagram $c'$ of Fig.~\ref{fig_red42222}. 
If $\delta=4$ there is a next-to-extremal quartic vertex at $z$. The
diagram gives zero since the $z$-integral is finite 
(Appendix) and the coupling
vanishes. If $\delta=6$ both vertices have vanishing couplings as they 
are extremal and the
integral is finite, so the diagram vanishes as well.

In diagram $d$ both $\delta=2,4$ are possible. In both cases the vertex at
$y$ is sub-extremal and the integral finite. If $\delta=2$, the quartic
coupling at $y$ is next-to-extremal and has a vanishing coupling. If
$\delta=4$ both vertices are extremal and have vanishing
couplings. Therefore this diagram does not contribute.

Let us now study the same set of diagrams when $SU(2,2|4)$
descendents are exchanged. We will show that descendent contributions
to diagrams $a-d$ of Fig.~\ref{fig_AdS42222} all vanish or give
corrections to the four-point structure in~(\ref{red42222a}). Since
the discussion is very technical, the reader interested only
in the main flow of the argument may wish to proceed to the summary
paragraph below.

$SU(4)$ flavour symmetry restricts the quantum number of internal
lines according to the Clebsch-Gordan decomposition ($q\leq k$)
\be
[0,k,0]\otimes [0,q,0] = \bigoplus_{\mu=0}^q
\bigoplus_{\nu=0}^{q-\mu} [\nu,k+q-2\mu-2\nu,\nu] \, .
\ee
In the present application $[0,k,0]$ and $[0,q,0]$ are external chiral
primary fields dual to operators of dimension $k$ and $q$,
respectively, while the exchanged fields in representations
$[\nu,j,\nu]$ on the right side are either primaries (for
$\nu=0$) of dimension $\Delta_\phi=j$ or descendents (for $\nu\geq
0$) of dimension $\Delta_\phi>j+2\nu$. A descendent must descend
from a chiral primary in the representation $[0,l,0]$ with
$\Delta_\phi > l \geq j+2\nu$.

We shall also use a fact which follows from the unique representation
of given three-point function as an extended superspace invariant. All
component bulk couplings $s_k s_q \phi$, $s_k s_q \phi^\prime$,
$s_k\phi\phi^\prime$ are then related by extended supersymmetry to
those of the scalar primaries in the same multiplet,
$s_k s_q \tilde{s}$, $s_ks_q\tilde{s}^\prime$,
$s_k\tilde{s}\tilde{s}^\prime$, where $\tilde{s}$ and
$\tilde{s}^\prime$ are the chiral primaries from which $\phi$ and
$\phi^\prime$ descend. In particular, descendent couplings vanish if
the associated primary vertex vanishes because of extremality. The
final link of our argument is the convergence or divergence of the AdS
integrals, as discussed in the Appendix.

We start with diagram $b$, where $SU(4)$ symmetry allows the
combinations of exchanged fields $\phi$, $\phi^\prime$ in the
following representations:
\be
\begin{array}{lllllll}
\phi = [0,2,0] \rightarrow & \phi^\prime = [0,0,0], & [0,2,0], &
[1,0,1], & [0,4,0], & [1,2,1], & [2,0,2] \\
\phi = [0,4,0] \rightarrow & \phi^\prime = [0,2,0], & [0,4,0], &
[1,2,1] &&& \\
\phi = [1,2,1] \rightarrow & \phi^\prime = [0,2,0], & [1,0,1], &
[0,4,0], & [1,2,1], & [2,0,2] & \\
\phi = [0,6,0] \rightarrow & \phi^\prime = [0,4,0] &&&&& \\
\phi = [1,4,1] \rightarrow & \phi^\prime = [0,4,0], & [1,2,1] &&&& \\
\phi = [2,2,2] \rightarrow & \phi^\prime = [0,4,0], & [2,0,2] &&&& 
\end{array} \label{repsb}
\ee
If $\phi$ is a chiral primary in the $[0,2,0]$, while $\phi^\prime$ is
any descendent, the exchange diagrams are infinite and made finite by
analytic continuation of the highest dimension $4\rightarrow 4-\eps$
both in the integrand and in the vanishing extremal coupling. The
limit $\eps \rightarrow 0$ reduces all diagrams to $b'$ of
Fig.~\ref{fig_red42222} and results in a correlation function of the
structure~(\ref{red42222a}), \ie, the product of the two-point
function $\la \O_2\O_2 \ra$ and a radiatively corrected four-point
function $\la \O_2\O_2 \O_2 \O_2\ra$. In all other cases in
(\ref{repsb}), $\phi$ has scale dimension $\Delta_\phi>2$, and the AdS
integral is finite. On the other hand at least one of the couplings
vanishes because the associated primary vertex is either extremal or
forbidden by $SU(4)$ symmetry.

For descendents in diagram $a$ of Fig.~\ref{fig_AdS42222}, the allowed
representations are
\be
\begin{array}{lllllll}
\phi= [0,0,0] \rightarrow &  \phi^\prime = [0,4,0] &&&&&  \\
\phi = [0,2,0] \rightarrow & \phi^\prime = [0,2,0], & [0,4,0], &
[1,2,1] &&& \\
\phi = [1,0,1] \rightarrow & \phi^\prime = [1,2,1], & [0,4,0] &&& \\
\phi = [0,4,0] \rightarrow & \phi^\prime = [0,0,0], & [0,2,0], &
[1,0,1], & [0,4,0], & [1,2,1], & [2,0,2] \\
\phi = [1,2,1] \rightarrow & \phi^\prime = [0,2,0], &
[1,0,1], & [0,4,0], & [1,2,1], & [2,0,2] & \\
\phi = [2,0,2] \rightarrow & \phi^\prime = [0,4,0], & [1,2,1], &
[2,0,2] &&& \\
\end{array} 
\ee
The case where $\phi$ and $\phi^\prime$ are chiral primaries in the
$[0,2,0]$ is the no-descendent case considered above. In all other
cases $\Delta_\phi+\Delta_{\phi^\prime}>4$ and the integrals are
finite, while at least one of the (associated) couplings is extremal
and vanishes. Hence there are no descendent contributions to diagram $a$.

In diagram $c$, the exchanged field $\phi$ can be in the
representations $[0,2,0]$, 
$[0,4,0]$, $[1,2,1]$, $[0,6,0]$, $[1,4,1]$ and $[2,2,2]$. In all cases
except the $[0,2,0]$ primary discussed above, we have descendent or
primary fields in excited Kaluza-Klein multiplets. The quartic
couplings $\phi s_2 s_2s_2$ to the chiral primary of the graviton
multiplet then vanish by consistent truncation. Since the integral is
finite, all of these contributions vanish.

In diagram $d$, $\phi$ can be in any of the representations $[0,0,0]$,
$[0,2,0]$, $[1,0,1]$, $[0,4,0]$, $[1,2,1]$ and $[2,0,2]$. Integrals
are always convergent. If $\phi$ comes from a chiral primary in the
$[0,\tilde{\delta},0]$ with $\tilde{\delta} \geq 3$, the coupling at
the lower 
vertex vanishes because the associated vertex of chiral primaries is
forbidden or extremal (or alternatively by consistent truncation). If,
on the other hand, $\tilde{\delta}=2$, $\phi$ is in the multiplet
of the graviton and consistent truncation forbids the upper vertex.

Summarizing, we have seen that the descendent-exchange diagrams only
contribute to the $\la \O_2\O_2\O_2\O_2 \ra$ factor in the
reduced diagram 
$b'$. Observe that diagrams $b'$ and $c'$ in
Fig.~\ref{fig_red42222} have 
the same factored structure as graphs $b$ and $b2$ in the SYM
calculation (Figs.~\ref{fig_free42222} and~\ref{fig_g42222}), whereas
diagram $a'$ in Fig.~\ref{fig_red42222} has the same structure as
graph $a$ in Fig.~\ref{fig_free42222}. Adding all the non-vanishing
contributions, which are given by $a'$, $b'$
and $c'$, we find
\bea
\lefteqn{\la \O_4(x) \O_2(x_1) \O_2(x_2) \O_2(x_3) \O_2(x_4)
\ra_{\rm exchange} }  && \nn &=&
\frac{4}{N}  
\, \la \O_2(x) \O_2(x_1)\O_2(x_2) \ra\, \la \O_2(x)\O_2(x_3)\O_2(x_4) 
\ra \nn
&& +
\frac{4}{N} \la \O_2(x) \O_{2}(x_1) \ra \, 
  {\la \O_2(x)\O_2(x_2)
\O_2(x_3)\O_2(x_4) \ra}_{\rm AdS}
 + \mbox{perms.}.  \label{facAdS42222} 
\eea 
$ {\la \O_2(x)\O_2(x_2)
\O_2(x_3)\O_2(x_4) \ra}_{\rm AdS}$
is the full AdS four-point correlator. It
includes exchange (diagram $b'$) and contact (diagram $c'$)
contributions. This function has been calculated very recently in
\cite{Frolov4222}. We note that the AdS and field theory four-point functions
cannot be compared directly since they are renormalized correlators at strong
and weak coupling respectively.
Nevertheless comparing Eqs. (\ref{facAdS42222})
and (\ref{fac42222}), we see that the same general space-time
structure is found in both the order-$g^2$ field theory calculation
and the exchange diagram contribution to the AdS calculation.

Finally, there is a contact contribution given by diagram $e$ of
Fig.~\ref{fig_AdS42222}:  
\bea
\lefteqn{\la \O_4(x) \O_{2}(x_1) \O_{2}(x_2) \O_{2}(x_3) 
\O_{2}(x_4) \ra_e} && \nn
 &\sim & {\cal G}(4,2,2,2,2) \int \! \frac{d^5 y}{y_0^5} K_4(x,y)
K_2(x_1,y) K_2(x_2,y) K_2(x_3,y) K_2(x_4,y).
\eea
As shown in the Appendix, this integral is finite. Furthermore it does
not have the factored form we have just 
discussed: it is not a product of a free function times another
factor. Hence the entire AdS calculation of this five-point function
does not give the same space-time structure we found at order
$g^2$ unless the coupling ${\cal G}(4,2,2,2,2)$ vanishes. On the other
hand this coupling does vanish according to consistent truncation.
Therefore if consistent
truncation holds, the space-time structure in the full AdS
calculation agrees with the field theory results. This, 
together with the results for extremal and next-to-extremal functions,
suggests that the factorization properties of
(sub)-extremal correlators may be extrapolated from weak to strong
coupling, at least in 
the large $N$ limit. This is our basic assumption in order to
derive the vanishing of certain supergravity couplings.


\section{General near-extremal correlators in field theory}
\label{CFT}

In this section we study near-extremal correlation functions in $\N=4$
SYM to order $g^2$ and show that all the Feynman graphs contributing
to an $E_n^m$ function factor into at least $n-m-1$ pieces which have
only one point in common: the point at which the highest-dimension
operator is inserted. In the following, ``factor'' (or ``piece'')
should be understood in that sense.  

First, we recall the results in \cite{Bianchi} and~\cite{EP}, where
it was shown that extremal and next-to-extremal $n$-point ($E_n^0$ and
$E_n^1$, respectively) functions
are not renormalized to order $g^2$. This means that these correlators
have a free-field form and, moreover, the overall coefficient does not
depend on $g$ . In the extremal case, each of the
non-vanishing Feynman graphs is a product
of $n-1$ (free) two-point functions. The point
at which the highest-dimension operator is inserted is common to all
the two-point functions. In the next-to-extremal case there is
in addition a propagator connecting two of the other operators, such
that the corresponding graphs are a product of $n-2$ factors: $n-3$
(free) two-point functions times one (free) next-to-extremal
three-point function. 

\begin{figure}[ht]
\begin{center}
\epsfxsize=10cm
\epsfbox{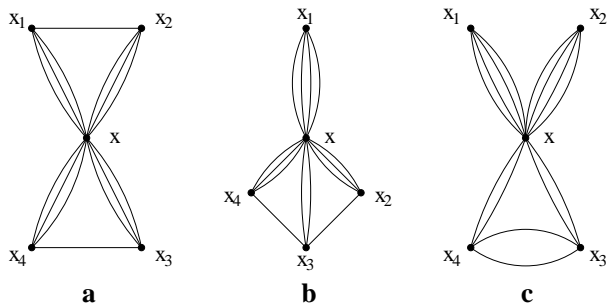}
\end{center}
\caption{Graphs contributing to $E_5^2$ functions at the
free-field level. Diagram $c$ is disconnected in the particular case
$k=4$, $k_i=2$ and therefore was not considered in Section 2.1.
\label{fig_freen2n2e}}  
\end{figure}
Next we consider a general $E_5^2$ function, $\la \O_{k}(x)
\O_{k_1}(x_1) \O_{k_2}(x_2) \O_{k_3}(x_3) \O_{k_4}(x_4) \ra$ with
$k=k_1+k_2+k_3+k_4-4$.  
The Feynman graphs contributing at the free-field level are shown in
Fig.~\ref{fig_freen2n2e}.

\begin{figure}[ht]
\begin{center}
\epsfxsize=13cm
\epsfbox{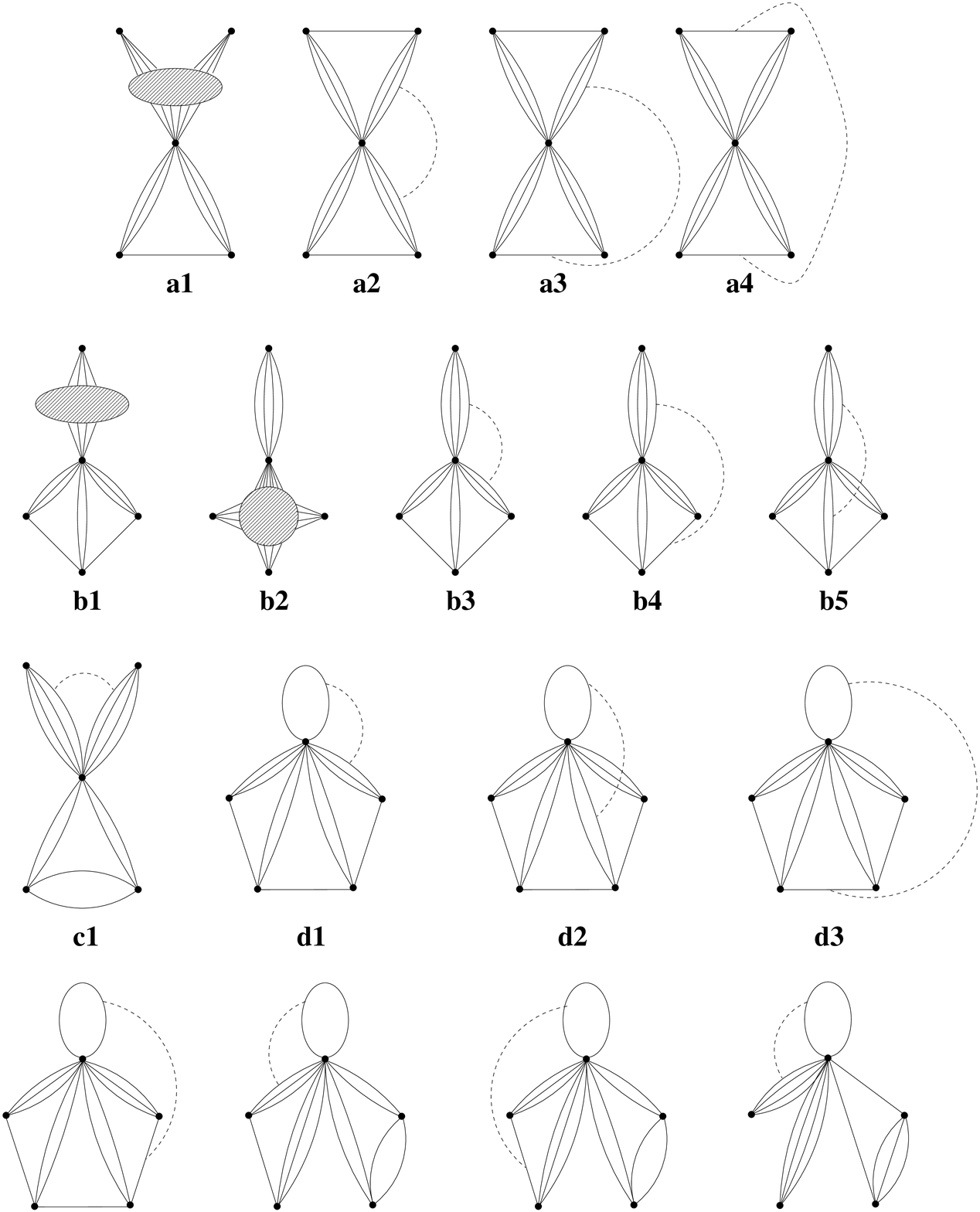}
\end{center}
\caption{Feynman graphs contributing to an $E_5^2$ function  at order
$g^2$. \label{fig_g2n2n2e}} 
\end{figure}
The Feynman graphs that contribute at order $g^2$ are depicted in
Fig.~\ref{fig_g2n2n2e}. The graphs $a1-c1$ are radiative corrections
of the graphs in Fig.~\ref{fig_freen2n2e}. Graphs
with two ``tadpoles'' connected by a 
dashed line vanish trivially due to the symmetry in the colour
indices in the highest-dimension operator, and we do not display them
explicitly here or in the subsequent. Graphs
$a1$ and $b1$ vanish due to non-renormalization theorems
for two-point and three-point correlators \cite{FreedmanCFT}. Indeed, as
was discussed in~\cite{EP}, the operator $\O_{k^\prime}(x)$ entering
the relevant two-point and three-point functions is in the
$[0,k^\prime,0]$ representation of the $SU(4)$ flavour group
and the extra colour-group generators appearing in the
trace of this operator do not spoil the validity of the
non-renormalization proofs
in~\cite{FreedmanCFT}. As in Section~2, all the remaining graphs
except $b2$ vanish due to the symmetries in the colour indices. We
show this explicitly for graph $b4$.

In order to simplify the equations we introduce the following
schematic notation:  
\be
\{ b_1 b_2 \cdots b_s\}  =  \Str (T^{b_1} T^{b_2} \cdots T^{b_s}
T^{a_i} \cdots T^{a_{i+r-1}}), 
\ee
where $a_i,\ldots,a_{i+r-1}$ are the colour indices of the $r$ legs of
a given operator which are attached to the highest-dimension operator,
$\O_k(x)$, and
$b_1,\ldots,b_{s}$ are extra colour indices. Moreover we write
the trace of the operator $\O_k(x)$ as
\be
\{a \} = \Str (T^{a_1} \cdots T^{a_{k}}),
\ee
and a trace with a commutator as, \eg,
\be
\{ [b,c] \} = \Str ([T^b,T^c] T^{a_i} \cdots T^{a_{i+r-1}}),
\ee
where the trace on the \rhs\ is not symmetrized under $b\leftrightarrow c$.
With this notation, the contribution of graph $b4$ reads
\bea
\lefteqn{b4 \sim \left(A f^{a_1bp}f^{cdp} + B f^{a_1cp}f^{bdp} + C f^{a_1dp}
f^{bcp}\right)} && \nn  
&& \mbox{} \cdot \; \Str(T^{a_1}\cdots T^{a_k}) Str(T^b T^{a_2}\cdot
T^{a_{k_1}}) \Str(T^d T^{a_{k_1+1}}\cdots T^{a_{k_1+k_2-1}}) \nn 
&& \mbox{} \cdot \; \Str(T^cT^e T^{a_{k_1+k_2}} \cdots
T^{a_{k_1+k_2+k_3-3}}) \Str(T^e T^{a_{k_1+k_2+k_3-2}} \cdots
T^{a_{k_1+k_2+k_3+k_4-4}}) \nn
& \equiv & \left(A f^{a_1bp}f^{cdp} + B f^{a_1cp}f^{bdp} + C f^{a_1dp}
f^{bcp}\right) \nn
&& \mbox{} \cdot \; \{ a \} \{b\} \{d\} \{ce\} \{e\},
\eea
where $A$, $B$ and $C$ are functions of space-time and flavour. We
show in turn that the three terms vanish, using the identities
$[T^a,T^c]=if^{acp} T^p$ and
\be
\sum_{i=1}^r \Tr(M_1\cdots [M_i,N] \cdots M_r)=0, \label{traceid}
\ee
where $M_i$, $N$ is any set of matrices. The same identity holds for
the symmetric trace.

For the term with coefficient $A$, contracting one of the structure
constants with a group constant inside a trace, we find
\bea
\mbox{$A$-term} & \sim & \{a\} \{ [a_1,p] \} \{d\} \{ce\} \{e\} \nn
&\sim& \{a\} \sum_i \{p[a_1,a_i]\} \{d\} \{ce\} \{e\} \nn 
&=& 0,
\eea
where in the last identity we have used the fact that in each term
there is a
contraction of an antisymmetric commutator, $[T^{a_1},T^{a_i}]$, 
times a symmetric trace, $\{ a \}$.
Similarly, for the term with coefficient $B$ we have
\bea
\mbox{$B$-term} & \sim & \{a\} \{b\} \{d\} \{[a_1,p]e\}\{e\} \nn
&\sim & \{a\} \{b\} \{d\} (\, \{p[a_1,e]\} + \sum_i \{pe[a_1,a_i] \,)
\{e\} \nn &\sim & \{a\} \{b\} \{d\} \{pq\} f^{a_1eq} \{e\} \nn
&\sim & \{a\} \{b\} \{d\} \{pq\} \{[a_1,q]\} \nn
&\sim & \{a\} \{b\} \{d\} \{pq\} \sum_i \{q[a_1,a_i]\} \nn
&=& 0.
\eea
Finally, let us show that the term with coefficient $C$ vanishes as
well:  
\bea
\mbox{$C$-term} & \sim & \{a\} \{b\} \{[a_1,p]\}\{ce\}\{e\} \nn
& \sim & \{a\} \{b\} \sum_i \{p[a_1,a_i]\} \nn
&=& 0.
\eea
Hence graph $b4$ vanishes. The essential step in the proof is
to use the identity (\ref{traceid}) to shift antisymmetric
combinations of indices arising from the structure constants between
traces, until a commutator $[T^{a_i},T^{a_j}]$ is obtained, which
vanishes when contracted with $\{a\}$. One can prove in a similar
manner that graphs $a2-a4$, $b3$, $b5$, $c1$, and $d1-d7$
vanish as well.   
On the other hand, there are graphs of the general
form $b2$ that give finite contributions because they cannot be
reduced to a sum 
of terms containing a conmutator $[T^{a_i},T^{a_j}]$. Nevertheless we
see that $b2$ has the factored form
\bea \lefteqn{
\la \O_k(x)\O_{k_1}(x_1)\O_{k_2}(x_2)\O_{k_3}(x_3)\O_{k_4}(x_4)\ra_{b2}}
& & \nn & =& \frac{1}{N} \sqrt{ k k_1 (k-k_1)}
 {\la \O_{k_1}(x)\O_{k_1}(x_1) \ra}  {\la
\O_{k-k_1}(x)\O_{k_2}(x_2)
\O_{k_3}(x_3)\O_{k_4}(x_4) \ra}^{(1)},
\eea
which is a product of a free two-point function and an $E_4^2$
four-point function. In this equation it is understood that the
colour generators of the operators $\O_{k_1}(x)$ and $\O_{k-k_1}(x)$
are included inside the same trace, such that the two factors are
coupled by colour. However, the colour indices within the first factor
are contracted and eventually a product of two functions
which are colour singlets is obtained. Therefore all the
graphs factor into at least two pieces and
the $E_5^2$ function has, to order $g^2$, the form
\bea
\lefteqn{\la \O_{k}(x) \O_{k_1}(x_1) \O_{k_2}(x_2) \O_{k_3}(x_3)
\O_{k_4}(x_4) \ra = \frac{1}{N} \sqrt{ k (k_1 + k_2-2)(k_3 + k_4-2)}} && 
\nn
&& \hspace{0.5cm} \cdot \;
\la \O_{k_1+k_2-2}(x) \O_{k_1}(x_1) \O_{k_2}(x_2) \ra
\la \O_{k_3+k_4-2}(x) \O_{k_3}(x_3) \O_{k_4}(x_4) \ra \nn 
&+& 
\frac{1}{N} \sqrt{ k k_1 k_2 (k_3+k_4-4)}
 {\la \O_{k_1}(x)\O_{k_1}(x_1) \ra} {\la \O_{k_2}(x)\O_{k_2}(x_2)
\ra}\la \O_{k_3+k_4-4}(x) \O_{k_3}(x_3) \O_{k_4}(x_4) \ra \nn  
&+&
\frac{1}{N} \sqrt{ k k_1 (k-k_1)}
 {\la \O_{k_1}(x)\O_{k_1}(x_1) \ra}  {\la
\O_{k-k_1}(x)\O_{k_2}(x_2) \O_{k_3}(x_3)\O_{k_4}(x_4) \ra}^\prime 
 + \mbox{perms.},  \label{face52} 
\eea
where
\be 
\la \O_{k_1+k_2-2p} (x
)\O_{k_1} (x_1)\O_{k_2} (x_2) \ra \, = \,
\frac{1}{N} \frac{ \sqrt{(k_1+k_2 -2p) k_1 k_2}}{
(x-x_1)^{2(k_1-p)} (x-x_2)^{2(k_2-p)} (x_1-x_2)^{2p}}  \, ,
\ee
and the square roots account for the large $N$ Wick factors and the
normalization. The prime in the four point function in~(\ref{face52})
indicates that a factorized free-field contribution to this four point
function has been explicitly written in the second term of the
equation and must not be included again.
Observe that the sum of the degrees
of extremality of the factors in each term is 2 (\ie,
next-to-next-to-extremal).
Schematically Eq.~(\ref{face52}) can be written
\be
E_5^2 = E_3^1 E_3^1 + E_2^0 E_2^0 E_3^2 + E_2^0 E_4^2.
\ee

The extension of this detailed analysis to $E_n^m$ functions for $n>5$
and $m\geq 
2$ is very tedious, and we limit the discussion to the demonstration
of factorization properties through order
$g^2$. Specifically we will argue that
\begin{itemize}

\item $E_6^2$ functions split into at least three factors:
\be E_6^2=E_2^0 E_3^1 E_3^1 + E_2^0 E_2^0 E_4^2 \, . 
\label{structureE62}
\ee 

\item $E_6^3$ functions split into at least two factors:
\be E_6^3=E_2^0 E_5^3 + E_3^1 E_4^2 +E_3^2E_4^2+E_3^3 E_4^0 \, .
\label{structureE63} \ee

\item In the general case $E_n^m$ splits into at least $n-m-1$
factors:
\be E_n^m= \sum_{\{n_j,m_j\}} \prod_{i=1}^{n-m-1} E_{n_i}^{m_i},
\label{genfac}
\ee
where $\sum_{i=1}^{n-m-1} n_i=2(n-1)-m$ and
$\sum_{i=1}^{n-m-1} m_i=m$. If $m\leq n-3$ each term
factors into at least two 
pieces and the maximum value of $n_i$ is $n-1$.

\end{itemize}
The qualification ``at least'' is meant to indicate that there are
graphs in which more factors occur. For instance, in graph $c$ of
Fig.~\ref{fig_e62} below, the factor $E_4^2$ of the last term in
(\ref{structureE62}) splits in turn into $E_2^0 E_3^2$. Even with
this limited 
aim, the discussion is complicated and some readers may wish to
proceed to Section~4.

\begin{figure}[ht]
\begin{center}
\epsfxsize=12cm
\epsfbox{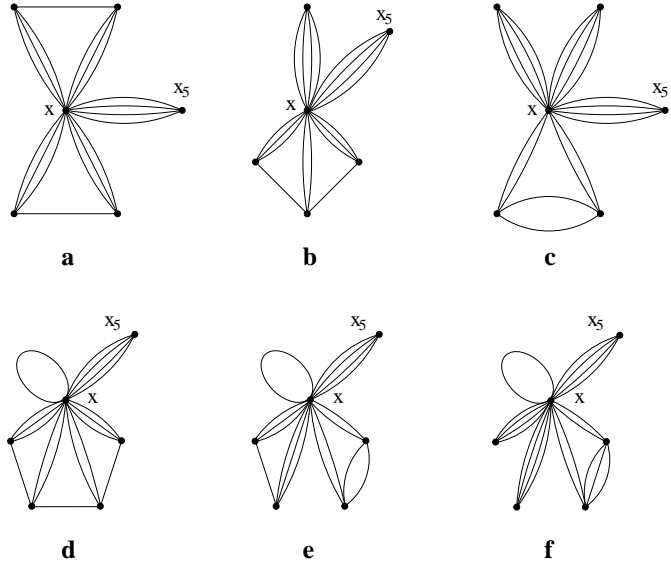}
\end{center}
\caption{Feynman graphs contributing to an $E_6^2$ function at the
free-field level.
\label{fig_e62}} \end{figure}
Let us consider six-point functions $\la \O_k(x)
\O_{k_1}(x_1) \cdots \O_{k_5}(x_5) \ra$. The $E_6^2$ correlators have
$k=\sum_{i=1}^5 k_i -4$. The free-field
Feynman graphs for such a function are displayed in
Figure~\ref{fig_e62}. We include graphs with tadpoles, which vanish
at this order (because the primary operators are traceless tensors in
flavour space) but can be used to construct order $g^2$
graphs. The graphs in Figure~\ref{fig_e62} are similar to the ones
contributing to the $E_5^2$ functions studied above, but with an
additional ``rainbow'' of propagators connecting the highest-dimension
operator to the additional operator (at $x_5$ in the figures). The
order $g^2$ corrections can be classified in three groups: graphs that
factor into the $E_5^2$-graphs times a free two-point subgraph
involving the new operator, corrections inside the rainbow and
graphs that connect the new 
rainbow to the rest of the free-field graph.  Using the results
above it is clear that the graphs of the first group have at
least three 
factors, two of which are free-field functions. The corrections within
the rainbow are corrections to a two-point function and
vanish. Finally, with the method used above for graph $b4$ of
Fig.~\ref{fig_g2n2n2e}, the
graphs of the third kind can be shown to 
either vanish or factor into three factors, two of which are
free-field functions. The second 
possibility occurs only when the new rainbow is connected to one of
the two lines at the bottom of graph $c$. Summarizing, all graphs
have at least three factors and we find the structure
in~(\ref{structureE62}). 

\begin{figure}[ht]
\begin{center}
\epsfxsize=12cm
\epsfbox{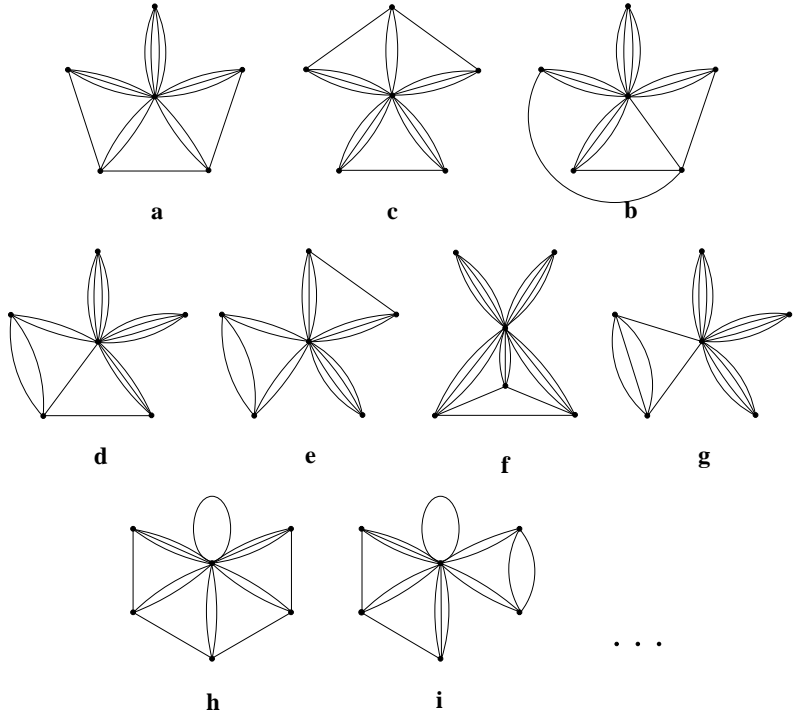}
\end{center}
\caption{Feynman graphs contributing to an $E_6^3$ function at the
free-field level. The dots stand for other graphs with one tadpole and
two or more pieces. \label{fig_e63}}
\end{figure}
The Feynman graphs that contribute at the free-field level to an
$E_6^3$ function ($k=\sum_{i=1}^5 k_i -6$) are depicted in
Figure~\ref{fig_e63}. It is useful to introduce the following
definition: We denote as
``T graphs'' those graphs which have no closed loops after
removing all the lines connected to the highest-dimension operator. A
T$_k$ graph is defined to be a
T graph with $k$ factors.
We can then distinguish four kinds of graphs in Figure~\ref{fig_e63}: 
\begin{enumerate}
\item Graphs with $q\geq 3$ factors: $d$, $e$, $f$ and $g$.
\item Graphs with one tadpole and at least two factors: $i$.
\item T$_2$ graphs: $a$, $b$ and $c$.
\item T$_1$ graphs with one tadpole: $h$.
\end{enumerate}
Let us consider now the order $g^2$ graphs, that are either
self-energy corrections or can be constructed inserting a dashed line
in the graphs of Figure~\ref{fig_e63}. Self-energy corrections do not
change the number of factors of the graph and we do not need to study
them, although we observe that they cancel other corrections in such a
way that two- and three-point factors are not
renormalized. From now on, we consider only dashed-line
graphs. Order $g^2$ graphs constructed by inserting a
dashed line into 
the free-field graphs of the first kind obviously have at least two
factors, as a single line cannot connect three or more pieces. The
same applies to order $g^2$ graphs constructed from graphs of the
second kind since one end of the dashed line has to be connected
to the tadpole line. Finally, we prove below for any number of points
that T graphs do not  
have order $g^2$ corrections connecting the different factors or one
factor to a tadpole. This means that order $g^2$ graphs constructed
from graphs of the third kind factor into two pieces and that the 
order $g^2$ graphs constructed from graphs of the fourth kind
vanish. Therefore we conclude that to order $g^2$ all non-vanishing
graphs factor into at least two pieces.

Furthermore each graph has a global degree of extremality equal to
3, \ie, it is a product $\prod_{i=1}^p E_{n_i}^{m_i}$ with
$\sum_{i=1}^p n_i= 6+p-1$ and $\sum_{i=1}^p m_i=3$. This property can
be read directly from the graphs in Figure~\ref{fig_e63}. 
Therefore we obtain the structure written
in~(\ref{structureE63}).

The analysis of the six-point correlators can be easily generalized to
$E_n^m$ functions, $\la \O_k(x) \O_{k_1}(x_1) \cdots
\O_{k_{n-1}}(x_{n-1}) \ra$ with $k=\sum_{i=1}^{n-1} k_i - 2m$. It
turns out that to order $g^2$ an $E_n^m$ 
function, $m\leq n-2$, is a sum of terms that factor into at least
$n-m-1$ pieces. The essential ingredient in the proof is the fact
that the Feynman graphs
contributing to an $E_n^m$ function at the 
free-field level fall among the following classes:
\begin{enumerate}
\item Graphs with $q\geq n-m$ factors.
\item Graphs with one tadpole and at least $n-m-1$ factors.
\item T$_{n-m-1}$ graphs.
\item T$_{n-m-2}$ graphs with one tadpole.
\end{enumerate}
To see this, consider first graphs without tadpoles. These
graphs have $m$ lines that are not connected to $x$. Let us
add these $m$ lines in turn. After adding one line we are left
with a graph with $n-2$ factors. Then we add another line. If it
connects the same two operators that were joined by the first line,
we obtain a non-T graph with $n-2$ factors. On the other hand, if
one end of the second line is attached to an operator that were not
connected to the first line, we get a T graph with $n-3$
factors. The third line can either connect different factors or
connect operators within one factor. In the first case, one gets a
non-T graph with $n-3$ factors or a T graph with $n-4$ factors,
depending on the choice of the second line. In the second case one
gets a non-T graph with $n-2$ or $n-3$ factors. We can proceed
recursively and find that after adding the $m$th line we get a T
graph with $n-m-1$ factors or a non-T graph with $q\geq n-m$
factors. If the graph has one tadpole there are $m+1$ lines that are
not connected to $x$, and the argument can be repeated with
$m\rightarrow m+1$.

The proof that the order $g^2$ graphs constructed from these ones
factor into ${n-m-1}$ graphs is identical to the proof
given above for $n=6$. The only non-trivial ingredient is that
order $g^2$ graphs constructed from free-field T$_{q}$ graphs
factor into $q$ parts or vanish. We show this next.

First, we observe that the following property holds:
\be
f^{bcp} \{c d_1 \cdots d_k\} \sim \sum_{i=1}^{r-1} f^{ba_iq} \{p d_1
\cdots d_k\} + \sum_{i=1}^k f^{bd_iq} \{p d_1
\cdots \widehat{d_i} \cdots d_k\} \, , \label{shifting}
\ee
where $r$ is the number of lines of the given operator connected to
the highest dimension operator. We see that all the terms on the \rhs\
have an $f$ with one of the indices of the $f$ on the \lhs, and one of the
indices in the trace at the \lhs. In other words: $p$ changes, $b$ is
replaced by $a_i$ or $d_i$ and $c$ stays the same.
\begin{figure}[ht]
\begin{center}
\epsfxsize=10cm
\epsfbox{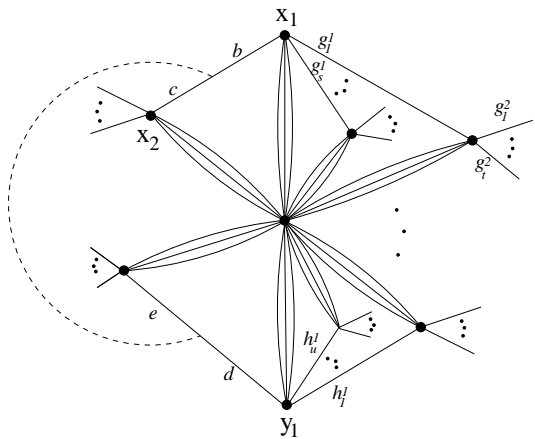}
\end{center}
\caption{Relevant part of a general order-$g^2$ correction to a T
graph. \label{fig_Tgraph}} 
\end{figure}

Consider now a general $T_p$ graph with $p\geq 2$ and one dashed line
connecting the ``external'' (\ie, not connected to the
highest-dimension operator) lines of two different factors. In
Fig.~\ref{fig_Tgraph} we show a part of this graph. There are three
contributions: 
\begin{enumerate}

\item
$f^{bcp} f^{dep}$. The index $b$ appears
in the trace of the operator at $x_1$. Using the relation
(\ref{shifting}) we obtain a sum of terms that contain $f^{a^1_icq^1}$ or
$f^{g^1_icq^1}$. This last $f$ is in turn
contracted with an index in another trace and we can apply the same
relation to obtain terms with $f^{a^2_icq^2}$ or with
$f^{g^2_icq^2}$. We repeat the procedure for the terms that do not
contain an $f$ with an $a_i$ until we reach an operator with only one
external leg. Then all terms contain $f^{a_icq}$. Since $c$
appears in the trace of the operator at $x_2$, we can use the same
method but keeping now $a_i$ fixed. Finally the expression is
reduced to a series of terms that contain $f^{a_ia_jq}$,
which is antisymmetric under $a_i\leftrightarrow a_j$. Since all
the terms also contain the symmetric trace $\{a\}$, they vanish.

\item
$f^{bdp} f^{cep}$. The indices in the first $f$ can be shifted just as
above until a sum of terms containing $f^{a_idq}$ is
obtained. This $f$ is contracted to the trace in $y_1$ through the
index $d$, and can be transformed into a sum of terms with either
$f^{a_i a^1_j q^1}$ or  $f^{a_ih^1_jq^1}$. The first kind of terms vanish when
contracted with $\{a\}$ and the second kind can be transform 
until all terms contain $f^{a_i a_j q^1}$. All these terms vanish when
contracted with $\{a\}$.

\item
$f^{bep} f^{cdp}$. The previous argument can be applied to this case
as well.

\end{enumerate}
The same can be argued if one or both ends of the dashed line were
connected to a rainbow or to a tadpole. Note that it is essential for
the argument 
that, when the indices in one $f$ are shifted by using
Eq.~(\ref{shifting}), the two new indices are different from all the
indices in previous steps. This would not occur if
there were loops of external lines. For this reason the proof only
applies to T graphs. Since the different factors of a T graph cannot
be connected at order $g^2$ we conclude that their factored structure is
preserve to this order.

Moreover, the sum of the degrees of extremality of the factors of any
term in an $E_n^m$ function is equal to $m$. Indeed, if a Feynman
graph factors into two or more functions it can be written as
$E_{n_1}^{m_1} E_{n_2}^{m_2}$, with $n_1+n_2=n+1$. In general, each
factor can be further 
factored. Both factors have $x$ as a common point. From the degrees
of extremality of the two factors it follows that
$\sum_{i=1}^{n_1-1} k_i - 2 m_1$ legs of the operator $\O_k(x)$ enter
the first factor and $\sum_{i=1}^{n_2-1} k_i - 2 m_2$ legs of this
operator enter the second factor. Since the operator $\O_k(x)$ has $k$
legs, 
\bea
k &=& \sum_{i=1}^{n_1-1} k_i - 2 m_1 + \sum_{i=1}^{n_2-1} k_i - 2 m_2
\nn 
&=& \sum_{i=1}^{n-1} k_i - 2 (m_1+m_2),
\eea
and from the $m$-extremality condition, $m=m_1+m_2$. The same procedure
can be applied to the subfactors of $E_{n_1}^{m_1}$ and
$E_{n_2}^{m_2}$. An alternative way of understanding this property is
noting that the degree of extremality of each factor is given by the
number of lines in that factor that are not connected to $x$. This
proves the structure written in~(\ref{genfac}).


\section{General near-extremal correlators in AdS}

\begin{figure}[ht]
\begin{center}
\epsfxsize=12cm
\epsfbox{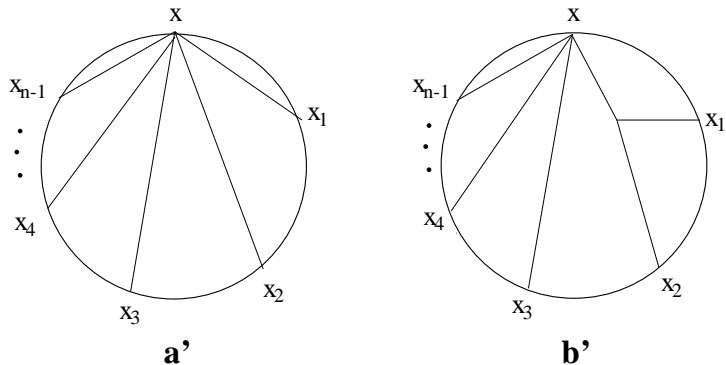}
\end{center}
\caption{Non-vanishing contributions to $E_n^0$
(diagram $a'$) and $E_n^1$ (diagram $b'$) functions.  
\label{fig_redn2e}}
\end{figure}

The $AdS_5\times S^5$ supergravity cubic and quartic couplings of
chiral primary fields have been calculated in \cite{Seiberg}
and~\cite{Frolov3}, respectively. In the extremal case, \ie, when the
conformal dimension of one field equals the sum of the remaining
conformal dimensions, these couplings vanish. Furthermore the
requirement that the AdS amplitudes have to be finite for non-coincident
points implies that the couplings of extremal $n$-field vertices must
vanish for any $n\geq 3$ \cite{Freedmanextremal}. As we show in the
Appendix, 
there are boundary contributions and the Witten diagrams of extremal
$n$-point functions reduce to a 
product of two-point functions connected to the point $x$ at which the
highest-dimension operator is inserted. This is
illustrated in Fig.~\ref{fig_redn2e}. 
The space-time structure of these diagrams agree with the free-field
approximation. Furthermore, in \cite{Freedmanextremal,West} it was
shown that the coefficient multiplying this structure is not
renormalized. 

In \cite{EP} it was shown that the next-to-extremal couplings of 
$n$ chiral primaries have to vanish, since otherwise
the AdS and the field-theoretical calculations of $E_n^1$ functions
would not agree. This has been recently checked
explictly for $n=4$ \cite{Frolovnew}. All non-vanishing Witten
diagrams contributing to the $E_n^1$ functions then reduce to products
of $n-3$
two-point functions times one next-to-extremal three-point function,
all of them with $x$ as a common point, as we show in
Fig.~\ref{fig_redn2e}. The space-time structure of the reduced diagrams
agrees with the field-theoretical calculation~\cite{EP}. For $E_4^1$
functions we have also checked that the coefficient of the AdS
structure agrees with the result found in field theory. The
calculation is similar to the ones we show in the Appendix below.

This analysis can be extended to other $E_n^m$ functions. It
can be shown in general that, as long as certain supergravity
couplings vanish, the AdS calculation gives the same space-time
structure as the order $g^2$ calculation, up to the detailed form of
the factors that are renormalized. We have already shown this for the
simplest $E^2_n$ function in
Section~\ref{subsec_AdS42222}.  

The calculation of a general $E_5^2$ function $\la
\O_k\O_{k_1}\O_{k_2}\O_{k_3}\O_{k_4}\ra$ in AdS is very similar to
the one for $\la \O_4\O_2\O_2\O_2\O_2\ra$ and we discuss only the new
features: 

1. We have
checked for general $k_i$'s that the coefficient of diagram $a$ in
Fig.~\ref{fig_AdS42222} agrees 
with the coefficient of the field-theory graph $a$ in 
Fig.~\ref{fig_freen2n2e} as given by the first term in
(\ref{face52}). This calculation may be found in the
Appendix below Eq.~(\ref{5app3}). 

2. Diagram $d$ of
Fig.~\ref{fig_AdS42222} does not vanish in general, but gives a new 
reduced diagram, $d'$ in Fig.~\ref{fig_red42222} (the same structure
results from diagram $b'$ when the upper vertex is extremal). This
reduced diagram 
has the same form as graph $c$ in Fig.~\ref{fig_freen2n2e}, which
gives the second  term in the field theory result (\ref{face52}). It is
interesting to note that this structure appears neither in
field theory nor in AdS for the connected $\la
\O_4\O_2\O_2\O_2\O_2\ra$. It is possible to calculate the
coefficient of diagram $d'$ as well, but this requires considerable
processing of the results for quartic couplings in~\cite{Frolov4}. 

3. The discussion of descendent exchange becomes more complicated.
For diagrams in which quartic couplings appear, consistent
Kaluza-Klein truncation is not sufficient for general $k_i$, but it
motivates the assumption\footnote{For scalar and vector descendents
there is evidence from AdS diagrams and field theory to support this
assumption. We do not discuss it here.} that a coupling with
descendents 
vanishes whenever the associated coupling of chiral
primaries does. With this assumption, one can easily
extend the analysis of descendents in Section~\ref{subsec_AdS42222}
and conclude that they only contribute to four-point factors. 

4. The contact diagram spoils the factorization property. Hence the result
in supergravity only agrees with the factorization we have found at
weak coupling if the $E_5^2$ couplings vanish. This is an
extension of consistent Kaluza-Klein truncation.

More general functions will be considered in the
next section. There we reverse the argument and derive the
vanishing of certain supergravity couplings from the requirement that
the field-theoretical calculation to order $g^2$ and the AdS
calculation give the same factored structure.


\section{Vanishing near-extremal supergravity couplings}
\label{sec_proof}

In this section we show that $E_n^m$ supergravity couplings vanish
for any $m\leq n-3$ if the following assumptions hold:
\begin{enumerate}
\item Witten diagrams are finite for non-coincident external points.
\item The factorization of $E_n^m$ functions, $m\leq n-3$ that we have
found (for any $N$) to order $g^2$ is preserved at large t'Hooft
coupling in the large $N$ limit.
\item A vertex with one or more descendents has a vanishing coupling
if the coupling of the associated chiral primary vertex vanishes.
\end{enumerate}
We do not use known results about supergravity couplings in the
argument. The fact 
that extremal cubic and quartic couplings and next-to-extremal quartic
couplings do vanish then supports the validity of the assumptions
above.

We also need the results of the Appendix, which are summarized by
\begin{itemize}
\item The integral over $y$ in a Witten diagram is
divergent if and only if the vertex at $y$ is extremal
and the highest-dimension field entering that vertex propagates into
the boundary (at point $x$).
\item If this is the case, the divergence is logarithmic and comes
from the region $y\sim x$. 
\end{itemize}
In analytic continuation this divergence
appears as a pole that, according to the first assumption, must be
cancelled by a zero coming from the coupling. When the regulator is
removed, the diagram factors into at least two pieces.

The proof proceeds by induction in the number of points, $n$. The
induction starts for $n=3$, $m=0$. We consider an extremal three-point
function. The integral is divergent and, according to the first
assumption, the coupling must vanish. Therefore, the $E_3^m$ couplings
vanish for any $m\leq 3-3=0$. Furthermore, the diagram factors into
two two-point functions, in agreement with the field-theoretical
results.

Let now $n\geq3$. Suppose that the $E_p^m$ couplings vanish for all
$m\leq p-3$, $p\leq n$. We have to show that the $E_{n+1}^m$ couplings
vanish for $m\leq 2$.

\begin{figure}[ht]
\begin{center}
\epsfxsize=15cm
\epsfbox{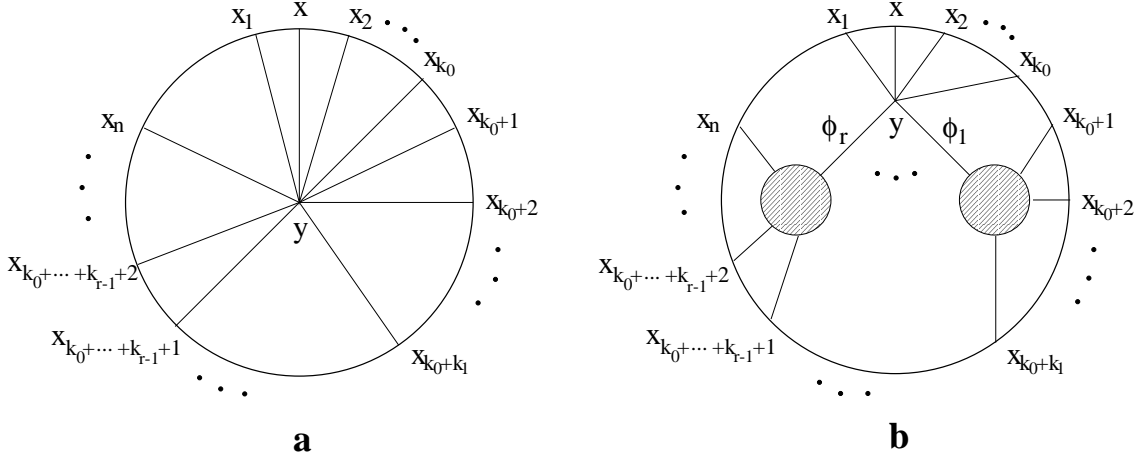}
\end{center}
\caption{Witten diagrams contributing to an $E_{n+1}^m$
function. \label{fig_contactexchange}} 
\end{figure}
Consider an $E_{n+1}^m$ function $\la \O_\Delta(x)
\O_{\Delta_1}(x_1)\cdots \O_{\Delta_n} \ra$,  with $m \leq
n-2$. This means that $\Delta=\sum_{i=1}^n \Delta_i - 2 m$. $\Delta$
and $\Delta_i$ are the conformal dimensions of the chiral primary
operators, which 
coincide with the non-vanishing Dynkin labels of their $SU(4)$
representation. Two kinds
of diagrams contribute: contact and exchange. All the exchange diagrams
are of the general form shown in Fig.~\ref{fig_contactexchange}, with
$1\leq 
k_0\leq n-2$. We consider
first diagrams that involve only primary exchanges. The exchanged
fields $\phi_1,\ldots,\phi_r$ are in the representations
$[0,\delta_1,0],\ldots,[0,\delta_r,0]$, and have dimensions
$\delta_1,\ldots,\delta_r$, respectively. Only the sets of fields with
\be
\sum_{j=1}^r \delta_j = \sum_{i=k_0+1}^n \Delta_i-2m, \sum_{i=k_0+1}^n
\Delta_i-(2m-1), \ldots, \sum_{i=k_0+1}^n \Delta_i.
\ee
are allowed by $SU(4)_R$.
We distinguish three possibilities:
\bea
&& 1.~~ \sum_{j=1}^r \delta_j= \sum_{i=k_0+1}^n \Delta_i-2m \, . \nn
&& 2.~~ \sum_{i=k_0+1}^n \Delta_i-2(m-1) \leq \sum_{j=1}^r \delta_j \leq
\sum_{i=k_0+1}^n \Delta_i -2(m-k-r+2) \, . \nn 
&& 3.~~ \sum_{j=1}^r \delta_j \geq \sum_{i=k_0+1}^n \Delta_i
-2(m-k_0-r+1) \, . \nonumber
\eea
\begin{figure}[ht]
\begin{center}
\epsfxsize=9cm
\epsfbox{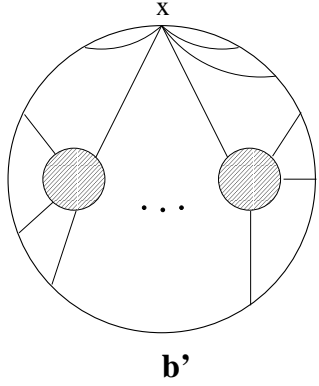}
\end{center}
\caption{Non-vanishing contribution to an $E_{n+1}^m$ function.
\label{fig_redexchange}}
\end{figure}
We study the three cases in turn.

{\em Case (1):\/} Since
$\sum_{i=k_0+1}^n \Delta_i-2m = \Delta-\sum_{i=1}^{k_0} \Delta_i$, the
vertex at $y$ is extremal and vanishes due to the induction hypothesis. 
On the other hand, the integral diverges since the field with highest
dimension, $\Delta$, propagates to the boundary. Hence the diagram
factors into ($E_2^0)^{k_0} E_{n-k_0+1}^m$, as shown in
Fig.~\ref{fig_redexchange}.  

{\em Case (2):\/} Since $\sum_{i=k_0+1}^n \Delta_i
-2(m-k-r+2)= \Delta - \sum_{i=1}^{k_0} \Delta_i +2 (k-1)$, the vertex
at $y$ is $E_{k_0+2}^p$ with $p\leq (k+2)-3$ and is zero by the
induction assumption. Because the integral over $y$ is convergent and
the other possible divergences are cancelled by zeros in the extremal
couplings, the diagram vanishes (see the Appendix).

{\em Case (3)} We have the inequality
\be 
\sum_{i=k_0+1}^n \Delta_i -2(m-k_0-r+1) \geq
\sum_{j=1}^r \left[ \sum_{i=k_0+\cdots + k_j+1}^{k_0+\cdots +k_{j+1}}
\Delta_i -2 (k_j-1)\right] +2.
\ee
If $\delta_1 \geq \sum_{i=k_0+1}^{k_0+k_1} -2(k_1-2)$, the subdiagram
involving the points $y$ and $x_{k_0+1}, \dots, x_{k_0+k_1}$ and the
points in the corresponding blob is an $E_{k_1+1}^p$ diagram with
$p\leq(k_1+1)-3$ and the highest-dimension field ($\phi_1$)
propagating into 
the bulk, which we denote as $\tilde{E}_{k_1+1}^p$. 
If, on the contrary, $\delta_1 \leq \sum_{i=k_0+1}^{k_0+k_1} -2(k_1-1)$,
the following relation holds:
\be 
\sum_{j=2}^r \delta_j  \geq
\sum_{j=2}^r \left[ \sum_{i=k_0+\cdots + k_j+1}^{k_0+\cdots +k_{j+1}}
\Delta_i -2 (k_j-1)\right] +2.
\ee
If $\delta_2 \geq \sum_{i=k_0+k_1+1}^{k_0+k_1+k_2} -2(k_2-2)$, the
corresponding subdiagram is an $\tilde{E}_{k_2+1}^p$ function with
$p\leq(k_2+1)-3$. Otherwise, we have a relation similar to the one
above involving $\sum_{j=3}^r \delta_j$. Iteratively, if 
$\delta_j \leq \sum_{i=k_0+\cdots+k_{j-1}+1}^{k_0+\cdots+k_j}
-2(k_j-1)$ for $j=1,\ldots,r-1$, we find that
\be 
\delta_r  \geq
\sum_{i=k_0+\cdots + k_{r-1}+1}^{k_0+\cdots +k_r}
\Delta_i -2 (k_r-2),
\ee
such that the corresponding subdiagram is an $\tilde{E}_{k_r+1}^p$
function with $p\leq (k_r+1)-3$.
Hence, at least one of the subdiagrams connected to $y$ is an
$\tilde{E}_{k_j+1}^p$ function with $p\leq (k_j+1)-3$.
The analysis of the entire diagram $b$ can be applied to this
subdiagram to show that it vanishes. We have
again three possibilities. The only difference is that in the
equivalent of case~(1) the integral finite because the
highest-dimension field ($\phi$) propagates into the bulk. More
details are given in the Appendix. The
iteration ends when all the subdiagrams are contact diagrams, which
vanish because the induction hypothesis implies that the $E_{k_j+1}^p$
coupling is zero for $p\leq (k_j+1)-3$. 

Therefore, the cases {\it (2)\/} and~{\it (3)\/} lead to vanishing
diagrams, whereas 
the case~{\it (1)\/} gives a reduced diagram with at least two
factors. Consider now the case when at least one of the fields
$\phi_j$ is a descendent. Then, if $\Delta_{\phi_j}$ denotes the
conformal dimension of each field, $\sum_{j=1}^r \Delta_{\phi_j} >
\sum_{i=k_0+1}^n \Delta_i-2m$ and the vertex at $y$ is
sub-extremal. Hence the integral over $y$ is finite. On the other hand,
if $\phi_j$ is a descendent, it descends from a chiral primary field
in a representation $[0,\tilde{\delta}_j,0]$ with $\tilde{\delta}_j >
(\delta_j)_{\rm min}$, where $(\delta_j)_{\rm min}$ is the minimal
dimension allowed by $SU(4)$ if $\phi_j$ were chiral primary. All the
possibilities can be discussed just as in the pure chiral primary
case, but changing some $\delta_j$ by $\tilde{\delta}_j$. Using
assumption~3, we conclude that in all cases
one or more of the couplings involved in the diagram vanish if the
induction hypothesis holds. 

We have shown that all the exchange diagrams either vanish or
factor into at least two pieces.
Finally, there is a contact diagram contributing to the $E_{n+1}^m$
function (diagram $a$ in Fig.~\ref{fig_contactexchange}). In the
extremal case, $m=0$, this diagram has a logarithmic 
divergence that has to be cancelled by a zero in the coupling. Hence
the $E_{n+1}^0$ 
coupling vanishes. If $m\geq 1$ the diagram is
finite and does not have a factored structure. Since the
non-vanishing exchange diagrams factor, the non-factored
contribution of the contact diagram survives in the full function. On
the other hand, we have shown in Section~\ref{CFT} that, to order
$g^2$, $E_{n+1}^m$ functions factor into at least two pieces if
$m\leq n-3$. According to the second assumption (and the Maldacena
conjecture), the AdS calculation should also give a factored
structure. Therefore, the $E_{n+1}^m$ couplings must vanish in order for
the contact diagram not to contribute. This completes the induction.

Finally, we observe that the first assumption can
be relaxed: It is sufficient 
to require that the sum of the Witten diagrams contributing to an AdS
amplitude is finite. This however, complicates the description of the
proof and for that reason we have chosen here an slightly stronger
assumption.


\section{Conclusions}

The main results of this paper are
\begin{enumerate}

\item Through order $g^2$, field theory graphs for near-extremal
$n$-point functions have a factored structure.

\item This structure is shown to be matched exactly by exchange
diagrams in AdS supergravity for the five-point function $\la
\O_4\O_2\O_2\O_2\O_2 \ra$. For more general $E_5^2$ functions an
additional technical assumption about descendent couplings is required
to reach the same conclusion.

\item For $E_5^2$ functions we show that there are unique supergravity
diagrams which match the corresponding field theory graphs in form and
coefficient. 

\item For $E_n^m$ functions with $n>5$ and $m\leq n-3$ we establish a
similar but less precise correspondence between field theory and
supergravity. 

\item The correspondence would be spoiled by non-factored supergravity
contact diagrams unless the associated near-extremal couplings
vanish. We therefore conjecture that $E_n^m$ couplings do vanish when
$m\leq n-3$. For some cases this is a consequence of consistent
Kaluza-Klein truncation and it is a natural extension of known results
for $n=3,4$. It remains to explain the pattern of vanishing couplings,
which is presumably a consequence of the reduction of Type II B
supergravity on the internal space $S^5$.

\end{enumerate}
\vspace*{1cm}

\subsection*{Acknowledgements}

We thank Leonardo Rastelli for useful discussions. The research of
E.D'H is supported in part by NSF Grants No. PHY-95-31023 and
PHY-98-19686 and the research of D.Z.F by NSF Grant No. PHY-97-22072.
J.E., who is a DFG Emmy Noether fellow, acknowledges funding through a 
DAAD postdoctoral fellowship. M.P.V. thanks MEC for a postdoctoral
fellowship. 


\section*{Appendix: AdS integrals}

\renewcommand{\theequation}{A.\arabic{equation}}

Here we study the integrals that appear in the calculation of
near-extremal correlation functions  
via the AdS/CFT correspondence. We first give some explicit examples
of integrals that appear in $E_5^2$ functions and then move to general
properties of $E_n^m$ integrals. We use the methods
developed in \cite{Freedman3pt}, \cite{Freedmanscalar4pt}, \cite{wrt} for
general correlators and in \cite{Freedmanextremal} and \cite{EP} for
extremal and next-to-extremal correlators. 
We use the Euclidean continuation of $AdS_5$ whose metric is
given by
\begin{equation}
ds^2 = \frac{1}{{z_0}^2} ( d{z_0}^2 + \sum\limits_{i=1}^{4} d{z_i}^2 )
\, .
\end{equation}
The scalar bulk to boundary propagator is given by \cite{Witten},
\cite{Freedman3pt} 
\bea
K_\Delta (x,z) = C_\Delta \left( \frac{z_0}{{z_0}^2
+ (\vec{z} - \vec{x})^2} \right)^\Delta  , \;&& C_\Delta = 
\frac{\Gamma(\Delta)} {\pi^2 \Gamma(\Delta - 2)}, ~ \Delta>2 \, , \nn
&& C_2 = \frac{1}{2\pi^2} \, .
\eea
$\Delta\geq 2$ is a real number.
Divergences in AdS integrals can arise only when the integration
points approach the boundary. Hence we need the behaviour of the
propagators when the bulk point approaches the boundary.
When $z_0\rightarrow 0$ but $\vec{z}\not \rightarrow \vec{x}$,
\begin{equation}
K_\Delta (x,z) \rightarrow C_\Delta z_0^\Delta \frac{1}{(\vec{z} -
\vec{x})^{2\Delta}}. \label{noncoincident}
\end{equation}
On the other
hand, when $z\rightarrow x$, 
\begin{equation}
K_\Delta (x,z) \rightarrow z_0^{4-\Delta} \delta(\vec{z}-\vec{x})\, .
\label{coincident}
\end{equation}

The explicit form of the bulk propagator $G_\delta(y,z)$ is not needed
here. Its behaviour 
when one of the points approaches the boundary
is~\cite{Freedmanscalar4pt} \cite{Rastelli}, in the
conventions of \cite{Seiberg},  
\begin{eqnarray}
G_{\delta} (y,z) \stackrel{y_0 \rightarrow 0}{\longrightarrow} 
 \tilde C_\delta \, 
{y_0}^\delta K_\delta(\vec{y},z) \, ,   && 
\tilde C_\delta = \frac{2^{\delta-6}}{ \delta -2} \frac{(2\pi)^5}{4 N^2}
\frac{(\delta+1)^2}{\delta (\delta-1)}, ~ k>2 \, , \nn
&& \tilde C_2 = \frac{(2\pi)^5}{4N^2} \frac{9}{2^5} \, .
\label{redprop}
\end{eqnarray}

\subsection*{Examples of five-point integrals}

We first illustrate the general properties
of AdS integrals by studying the contribution of diagram $b$ of
Fig.~\ref{fig_AdS42222} 
to a general next-to-next-to-extremal five-point function. We always
assume that the five points are non-coincident.
The conformal dimensions of the chiral primary
operators involved satisfy $\Delta = \sum_{i=1}^4 \Delta_i - 4$. We
take $\phi$ and 
$\phi^\prime$ to be chiral primaries of conformal dimension $\delta$
and $\delta^\prime$, respectively. The contribution of the diagram is
\bea \lefteqn{
\la \O_\Delta (x)\O_{\Delta_1} (x_1)
\O_{\Delta_2} (x_2)\O_{\Delta_3} (x_3)\O_{\Delta_4} (x_4)\ra } && \nn
&& = \frac{\G_3(\Delta,\Delta_1,\delta)
\G_3(\delta,\Delta_2,\delta^\prime) 
\G_3(\Delta_3,\Delta_4,\delta^\prime)}{\N_{\Delta}
\N_{{\Delta_1}} \N_{{\Delta_2}} 
\N_{{\Delta_3}} \N_{{\Delta_4}}}
\int \!\!\!\int \!\!\!\int  \, \frac{ d^5y}{y_0^5}\frac
{ d^5z}{z_0^5} \frac{ d^5w}{w_0^5} \;   K_\Delta(x,y)
K_{\Delta_1}(x_1,y)  \nn 
&&  \hspace{1cm} \cdot \;
G_\delta (y,z) 
K_{\Delta_2}(x_2,z) G_{\delta'}(z,w) K_{\Delta_3} (x_3,w)
K_{\Delta_4} (x_4,w) \, , \label{Ia} 
\eea
with $\N_{\Delta}$ as in (\ref{AdSnorm}). 
The possible divergence can only arise from the region $y\sim
x$. Indeed, in this region we can use 
Eqs.~(\ref{noncoincident}), (\ref{coincident}) and~(\ref{redprop})
to see that the integrand is proportional to 
\bea
{y_0}^{-5} {y_0}^{4-\Delta} {y_0}^{\Delta_1}
{y_0}^\delta & = & {y_0}^{\delta+\Delta_1-\Delta-1} \nn
& \equiv & y_0^{-\alpha-1}
 \label{counting} 
\eea
The degree of divergence of the integral over $y_0$ is given by
$\alpha$. The $SU(4)$ R-symmetry implies that $\delta
\geq \Delta - \Delta_1$ and thus $\alpha \leq 0$. Hence there are
two possibilites: the integral is finite if $\alpha<0$ and it is
logarithmically divergent if $\alpha=0$. Since $\Delta > \Delta_i$,
$i=1,\ldots,4$, a similar counting shows that the regions when one or
more of the integration points approach any $x_i$ do not give any
divergence. 

Let us consider first the divergent case which occurs when $\delta
= \Delta - \Delta_1$. We regularize the integral by analytic
continuation in the highest dimension which we write as
$\Delta-\eps$, $\eps>0$. The coupling of the vertex at $y$ has to
be changed accordingly. The integral is convergent since
now we have $\alpha=-\eps$. For small $\eps$ the integral is
dominated by the contribution of the region $y\sim x$. In this region
we can use Eq.~(\ref{redprop}) and write Eq.~(\ref{Ia}) as
\bea
\lefteqn{ \la \O_\Delta (x)\O_{\Delta_1} (x_1
)\O_{\Delta_2} (x_2)\O_{\Delta_3} (x_3)\O_{\Delta_4} (x_4)\ra_1 }  &&
\nn 
&& =  I_1^y\cdot \la \O_{\Delta-\Delta_1} (x)
\O_{\Delta_2} (x_2)\O_{\Delta_3} (x_3)\O_{\Delta_4} (x_4)\ra_1 \, 
\label{5app1}
\eea
with
\bea
&&  I_1^y =  \G_3(\Delta-\eps,\Delta_1,\delta)
\frac{C_\Delta C_{\Delta_1} \tilde C_\delta}{{(x-x_1)}^{2 \Delta_1}} 
\frac{ \N_{{\Delta - \Delta_1}}}{\N_{{\Delta }}
\N_{{\Delta_1}}} \, \int_\R \, \frac{d^5y}{{y_0}^5} \, 
\frac{{y_0}^{2\Delta - \eps
}}{  ({y} - {x})^{ 2(\Delta-\eps)}} \, , \hspace{4cm}  \\ 
&& \la \O_{\Delta-\Delta_1} (x
)\O_{\Delta_2} (x_2)\O_{\Delta_3} (x_3)\O_{\Delta_4} (x_4)\ra_1
 \nn
&& \hspace{.5cm} =  (\N_{{\Delta-\Delta_1}}
\N_{{\Delta_2}}\N_{{\Delta_3}}\N_{{\Delta_4}})^{-1}
\G_3(\delta,\Delta_2,\delta^\prime)
\G_3(\Delta_3,\Delta_4,\delta^\prime) \nn 
&& \hspace{1.5cm} \cdot \;
\int \!\!\int  \, \frac{ d^5z}{z_0^5}
\frac{ d^5w}{w_0^5} \; K_\delta (x,z) 
K_{\Delta_2}(x_2,z) G_{\delta'}(z,w) 
 \; K_{\Delta_3} (x_3,w) K_{\Delta_4} (x_4,w) \, .
\eea
$\R$ is any neighbourhood of $x$ where the approximation
(\ref{redprop}) is valid.
The integral in $I_1^y$ leads to a pole in $\eps$, which is
cancelled by a factor of $\eps$ arising from
$\G_3(\Delta-\eps,\Delta_1,\delta)$. This is described in detail in
\cite{EP}. The result when $\eps \rightarrow 0$ is
\begin{equation}
I_1^y \, = \frac{1}{N} \sqrt{\Delta \Delta_1 (\Delta-\Delta_1)} \, \la 
\O_{\Delta_1}(x)\O_{\Delta_1}(x_1) \ra \, .
\end{equation}
On the other hand, $\la \O_{\Delta-\Delta_1}(x) \O_{\Delta_2}
(x_2)\O_{\Delta_3}(x_3)\O_{\Delta_4} (x_4)\ra_1$ has the form of an
exchange 
contribution to a four-point function 
which may be evaluated using the methods of \cite{wrt}. 
We see that the five point function contribution (\ref{5app1})
factors into a free-field two-point function times a 
four-point function contribution.

As an example of a convergent integral, let us consider the case when
the dimension of the exchanged fields is $\delta=\Delta-\Delta_1+2$
and $\delta^\prime=\Delta_3+\Delta_4-2$. The integral is convergent
since $\alpha=-2$. Indeed, in this simple case we can perform the
integral over $y$ and $w$ explicitly using the methods of~\cite{wrt}
and find 
\bea
\lefteqn{
\la \O_\Delta (x)\O_{\Delta_1} (x_1
)\O_{\Delta_2} (x_2)\O_{\Delta_3} (x_3)\O_{\Delta_4} (x_4)\ra_2} &&\nn
  &= & \G(\Delta,\Delta_1,\Delta-\Delta_1+2)
\G(\Delta-\Delta_1+2,\Delta_2,\Delta_3+\Delta_4-2)
\G(\Delta_3+\Delta_4-2,\Delta_3,\Delta_4)  \nn
&& \hspace{.5cm}  \cdot \; a_{\Delta_4-1} \, \frac{1}{(x_3-x_4)^{2}}     
\sum_{k= \Delta - \Delta_1 + 1}^{\Delta-1}
\, b_k \,  \frac{1}{(x-x_1)^{2(\Delta-k)}}  \nn
&& \hspace{.5cm} \cdot \;  \int \! \frac{ d^5y}{y_0^5}  \;
K_{k}(x,y) K_{\Delta_1 - \Delta +k}(x_1,y)  
K_{\Delta_2}(x_2,y) K_{\Delta_3-1}(x_3,y)
K_{\Delta_4-1}(x_4,y) \, ,
\eea
with finite coefficients $a_{\Delta_4 -1}$, $b_k$. The remaining
$y$-integral is finite for every term in the sum~\cite{complete4}.
Therefore, 
since the second coupling is extremal and therefore vanishes, the whole
contribution vanishes. 

As a final example for a five-point function contribution we consider
diagram $b$ of Fig.~\ref{fig_AdS42222},
\bea \lefteqn{
\la \O_\Delta (x)\O_{\Delta_1} (x_1
)\O_{\Delta_2} (x_2)\O_{\Delta_3} (x_3)\O_{\Delta_4} (x_4)\ra_3
} && \nn
&& = \frac{\G_3(\Delta,\delta,\delta') \G_3(\Delta_1,\Delta_2,\delta)
\G_3(\Delta_3,\Delta_4,\delta^\prime)}{\N_{\Delta}\N_{{\Delta_1}}
\N_{{\Delta_2}}\N_{{\Delta_3}} \N_{{\Delta_4}}}
\; \int \!\!\!\int \!\!\!\int  \, \frac{ d^5y}{y_0^5}\frac
{ d^5z}{z_0^5} \frac{ d^5w}{w_0^5} \;   K_\Delta(x,y)
G_\delta (y,z)  \nn 
&&  \hspace{1cm} \cdot \;
K_{\Delta_1}(x_1,z) K_{\Delta_2}(x_2,z) G_{\delta^\prime}(y,w)
K_{\Delta_3} (x_3,w) 
K_{\Delta_4} (x_4,w) \, , \label{I3}
\eea
when $\delta= \Delta_1+ \Delta_2 - 2$ and $\delta^\prime = \Delta_3+
\Delta_4 - 2$.  
Again the vertex at $y$ is extremal, and the dominant contribution arises when
$y \sim x$. For this case the limit (\ref{redprop})
applies to both bulk-to-bulk propagators, such that the integral factors
into 
\begin{eqnarray} \lefteqn{
\la \O_\Delta (x)\O_{\Delta_1} (x_1
)\O_{\Delta_2} (x_2)\O_{\Delta_3} (x_3)\O_{\Delta_4} (x_4)\ra_3 
} && \nn &=&  I_3^y\cdot \la \O_{\Delta_1+\Delta_1-2} (x
)\O_{\Delta_1} (x_1)\O_{\Delta_2} (x_2) \ra \,
\la \O_{\Delta_3+\Delta_4-2} (x
)\O_{\Delta_3} (x_3)\O_{\Delta_4} (x_4) \ra \, , 
\label{5app3}
\end{eqnarray}
where
\be
I^y_3 = \tilde C_{\Delta_1+\Delta_2-2} \tilde C_{\Delta_3+\Delta_4-2} 
 C_\Delta \frac{ \N_{{\Delta_1 + \Delta_2 -2}}
\N_{{\Delta_3+\Delta_4 -2 }}}{
\N_{{\Delta}}} \G(\Delta-\eps, \delta,  \delta')
\, \int_\R \, \frac{d^5y}{{y_0}^5} \, 
\frac{{y_0}^{2\Delta - \eps
}}{  ({y} - {x})^{ 2(\Delta-\eps)}}
\ee
and
\bea \lefteqn{
\la \O_{\Delta_1+\Delta_1-2} (x
)\O_{\Delta_1} (x_1)\O_{\Delta_2} (x_2) \ra} && \nn
&=& \frac{\G(\delta, \Delta_1, \Delta_2)}{ \N_{{\Delta_1 + \Delta_2
-2}}  \N_{{\Delta_1}} \N_{{\Delta_2 }}}      \, 
\int  \,  \frac{ d^5w}{w_0^5} \;   K_\delta(x,w)
K_{\Delta_1} (x_1,w) K_{\Delta_2}(x_2,w)  \nn
&=& \frac{1}{N} \frac{ \sqrt{(\Delta_1+\Delta_2 -2) \Delta_1 \Delta_2}}{
(x-x_1)^{2(\Delta_1-1)} (x-x_2)^{2(\Delta_2-1)} (x_1-x_2)^{2}}  \, ,
\eea
such that we obtain
\bea
\lefteqn{\la \O_\Delta (x)\O_{\Delta_1} (x_1
)\O_{\Delta_2} (x_2)\O_{\Delta_3} (x_3)\O_{\Delta_4} (x_4)\ra_3} && \nn
&=& \frac{1}{N} \sqrt{\Delta (\Delta_1+\Delta_2 -2)(\Delta_3+\Delta_4 -2)}
\la \O_{\Delta_1+\Delta_1-2} (x
)\O_{\Delta_1} (x_1)\O_{\Delta_2} (x_2) \ra \, \nn && \hspace{2cm} \cdot
\la \O_{\Delta_3+\Delta_4-2} (x
)\O_{\Delta_3} (x_3)\O_{\Delta_4} (x_4) \ra \, 
\eea
which agrees exactly with the corresponding contribution to the field
theory result (\ref{face52}).

\subsection*{General AdS integrals}

Let us now turn to studying general $E_{n+1}^m$ functions with $m \leq
n-2$ (and non-coincident points). This is relevant for the inductive
proof of Section~\ref{sec_proof}. For simplicity we work with
non-derivative couplings. Derivative couplings do not change the
leading degree of divergence of the integrals: if $y$ is the
integration variable, the derivatives $\partial/\partial y_0$ make the
integrand more singular as $y$ approach the boundary, but this effect
is exactly compensated by the powers of $y_0$ arising from the metric.
On the other hand, we
assume in the following that all the extremal couplings
vanish. More precisely, if analytic continuation is used, we assume
that the extremal couplings contains a zero in the regulator. This is
required in order to obtain non-divergent Witten diagrams.
We consider first the
contact diagram (diagram $a$ in Fig.~\ref{fig_contactexchange}) and regularize
a possible divergence by analitical continuation in the highest
dimension, $\Delta$. This does not affect
the result if the integral is convergent. Up to couplings the diagram
gives 
\begin{equation}
\I_a = \int \! \frac{d^5 y}{{y_0}^5} \, K_{\Delta-\eps}(x,y)
\prod_{i=1}^n  K_{\Delta_i}(x_i,y).
\end{equation}
As $y\rightarrow x$ the integrand is proportional to ${y_0}^{\sum_i^n
\Delta_i - (\Delta-\eps)-1}$, \ie, $\alpha=\Delta -\eps - \sum_i^n
\Delta_i$ for this diagram. Therefore, for $\eps = 0$ the
integral diverges logarithmically if the function is extremal and is
convergent in any sub-extremal case. Obviously the integrand is less
singular in the regions $y \sim x_i$, as the $m$-extremality condition
implies that $\Delta_i < \Delta$.
In the extremal case, $\Delta=\sum_i^n \Delta_i$, the integral gives
rise to a pole in $\eps$ arising from the region $y\sim x$:
\bea
\I_a & \sim & \int_\R \! \frac{d y_0}{{y_0}} \, {y_0}^\eps 
\prod_{i=1}^n  \frac{1}{(x-x_i)^{2 \Delta_i}} \; + \; R_a   \nn
& \sim & \frac{1}{\eps} \prod_{i=1}^n  \frac{1}{(x-x_i)^{2
\Delta_i}} \; + \; R_a^\prime \, ,
\eea
where $R_a$ and $R_a^\prime$ are analytic in $\eps$.
The pole is cancelled by a zero in the
extremal coupling. Hence, the analytic part $R^\prime$ is irrelevant
when $\eps \rightarrow 0$, and the diagram factors into $n$
two-point functions.

\begin{figure}[ht]
\begin{center}
\epsfxsize=11cm
\epsfbox{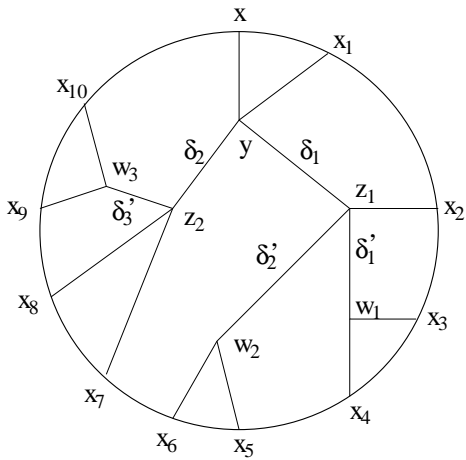}
\end{center}
\caption{Typical exchange diagram. \label{fig_typical}}
\end{figure}
A typical exchange diagram is shown in Fig.~\ref{fig_typical}. Let us
study the possible divergent limits. We consider first the case when
only one internal point, which we denote generically as $u$,
approaches a point $x^\prime$ in the 
boundary. If $x^\prime \not = x_i$, $i=0, \ldots, 9$ ($x_0\equiv x$)
or $x^\prime = x_i$ but $u$ is not directly connected to $x_i$ by a
bulk-to-boundary propagator, both
the bulk propagators and bulk-to-boundary propagators connected to $u$
are proportional to positive powers of $u_0$ as $u_0\rightarrow
0$. For example, if $y\rightarrow x_2$, the integrand
is proportional to ${y_0}^{\Delta+\Delta_1+\delta_1+\delta_2-5}$. In the
worst possible case the integrand is proportional to 
${u_0}^{2+2+2-5}$. Hence this region does not contribute to the
integral. On the other hand, if $x^\prime=x_i$ and $u$ is directly
connected to $x_i$ there is a power
${u_0}^{4-\Delta_i}$. For instance, if $y\rightarrow x$ the integrand
is proportional to ${y_0}^{\Delta_1+\delta_1+\delta_2-\Delta-1}$.
Therefore a logarithmic divergence appears in
this region if the vertex at $u$ is extremal and $\Delta_i$ is the
largest dimension entering this vertex. Otherwise the integral over
this region is finite. For instance, if $y\rightarrow x$ the integrand
is proportional to ${y_0}^{\Delta_1+\delta_1+\delta_2-\Delta-1}$. 

Let us now allow the possibility that two bulk points $u_1$ and $u_2$
approach the boundary. If they approach different points the two
limits can be considered independently. For instance, if $y\rightarrow
x$ and 
$z_1\rightarrow x_2$, the integrand is proportional to
${y_0}^{\Delta_1+\delta_1+\delta_2-\Delta-1}
{(z_1)_0}^{\delta_1+\delta_1^\prime+\delta_2^\prime-\Delta_2-1}$. If
$u_1$ and $u_2$ approach the same point $x^\prime$ and none of them is
directly connected to this point, all the propagators contribute with
positive powers of $(u_1)_0$ and $(u_2)_0$ and the double integral
over $u_1$ and $u_2$ is convergent. Finally, if $u_1\rightarrow x_i$
and $u_2\rightarrow x_i$ and $x_i$ is directly connected to, say,
$u_1$, there are two possibilities: If $u_1$ and $u_2$ are not
directly connected by a bulk propagator there is a factor of
${(u_1)_0}^{4-\Delta_i}$ but all the powers of $(u_2)_0$ are
positive. On the other hand, if $u_1$ and $u_2$ are directly connected
there are two factors with negative powers: 
${(u_1)_0}^{4-\Delta_i}{(u_2)_0}^{4-\delta_{12}}$, where $\delta_{12}$
is the dimension of the propagator connecting $u_1$ and $u_2$. For
example, if $y\rightarrow x$ and $z_2\rightarrow x$, the integrand is
proportional to ${y_0}^{\Delta_1+\delta_1+\delta_2-\Delta-1}
{(z_2)_0}^{\Delta_6+\Delta_7+\Delta_8^\prime-\delta_2-1}$. The
integral over $y$ ($z_2$) is then logarithmically divergent if the
vertex at $y$ ($z_2$) is extremal. In general, a (logarithmic)
divergence in the integral over an internal point $u$ can only
arise when the following two conditions are fulfilled: 
\begin{enumerate}
\item Either $u$ approaches a point $x_i$
that is connected to $u$ by a bulk-to-boundary propagator or it
approaches a point $x_i$ that is connected to $u$ by a 
string of propagators and all the points in those propagators also
approach $x_i$. 
\item The vertex at $u$ is extremal and the highest dimension of the
fields entering the vertex is the one of the field that connects $u$
(directly or indirectly) with $x_i$.
\end{enumerate}
We shall use these properties in the following. 

Any diagram with at least one exchange is of the general form
represented by diagram $b$ of Fig.~\ref{fig_contactexchange}, where we
have isolated the 
vertex at $y$ to which the highest-dimension operator is connected. 
This vertex involves
$(k_0+1)$ bulk-to-boundary propagators
and $r$ bulk-to-bulk propagators depending on $y$ and one $z_j$,
$j=1,\ldots,r$.  
There are $r$ subdiagrams involving further tree interactions depicted
by a shadowed circle. The $j$th subdiagram depends on the bulk
variable $z_j$ and on $k_j$ boundary variables
$x_{k_0 + \cdots + k_{j-1} + 1}, \ldots, x_{k_0 + \cdots + k_j}$ which
we collectively denote as $\tilde{x}_j$. 
We denote the contribution from the
$j$th subdiagram by $D_j(z_j, \tilde{x}_j)$.
Using analytic continuation,
the contribution of diagram $b$ is given by 
\begin{equation} 
\I_b =  \int \! \frac{d^5y}{y_0^5} K_{\Delta-\eps} (x,y) \, \prod_{i=1}^k 
K_{\Delta_i}(x_i,y)  \,
\prod_{j=1}^r   \int \! 
\frac{d^5 z_j}{{(z_j)_0}^5} \, G_{\delta_j}(y, z_j)
D_j(z_j, \tilde{x}_j) \, ,\label{I}
\end{equation}
where we have omitted the couplings. 
A divergence in the integral over $y$ can only arise from the region
$y \sim x$. The degree of divergence is given by
\begin{equation}
\alpha = \Delta - \sum_{i=1}^k \Delta_i - \sum_{j=1}^r
\delta_j -\eps \, .
\end{equation} 
Therefore the integral over $y$ is divergent (for $\eps=0$) if and
only if the vertex at $y$ is extremal, $\sum_{j=1}^r \delta_j = \Delta
- \sum_{i=1}^k \Delta_i$.
In this case ({\it case (1)} in the text), the integral is
dominated by 
the region $y \sim x$ which gives rise to a pole $1/ \eps$:
\bea
\I_b & \sim &  \int_\R \! \frac{d y_0}{y_0} {y_0}^\eps \,
\prod_{i=1}^k \frac{1}{(x-x_i)^{\Delta_i}}  \,
\prod_{j=1}^r   \int \! 
\frac{d^5 z_j}{{(z_j)_0}^5} \, K_{\delta_j}(x, z_j)
D_j(z_j, \tilde{x}_j) \; + \; R_b  \\
& \sim & \frac{1}{\eps} \prod_{i=1}^k \frac{1}{(x-x_i)^{\Delta_i}}  \,
\prod_{j=1}^r H_j(x,\tilde{x}_j)  \; + \; R_b^\prime \, ,
\label{generalfactor} 
\eea
where $R_b$ and $R_b^\prime$ are regular in $\eps$ and
\begin{equation}
H_j(x,\tilde{x}_j) \equiv  
\int \! \frac{d^5 z_j}{{(z_j)_0}^5} \, K_{\delta_j}(x, z_j)
D_j(z_j, \tilde{x}_j) \, .
\end{equation}
Since the extremal coupling contains a factor of $\eps$, the final 
result is finite and factors into $k$ two-point functions times a
product of $r$ functions. The $j$th of these functions has $k_j+1$
points.  This structure is illustrated in Fig.~\ref{fig_redexchange}.

Consider now the case when the vertex at $y$ is sub-extremal:
$\sum_{j=1}^r \delta_j > \Delta - \sum_{i=1}^k \Delta_i$. 
Then we have $\alpha<0$ (we set $\eps=0$ now) and the integral
over $y$ is convergent. We also have to study the behaviour of
the integrand when some of the other internal points approach the
boundary. As we have shown above, (logarithmic) divergences only arise
from extremal vertices. They manifest as poles in analytic
continuation and are cancelled by the zeros in the corresponding
extremal couplings. The diagram is then finite even if the
coupling of the vertex at $y$ is not included. Therefore if this
coupling vanishes, as in {\it case (2)} in the text, the diagram gives
zero. Finally, for {\it case (3)} in the text we need to show
that the diagram vanishes when one of the subdiagrams is an
$\tilde{E}_{k_j+1}^p$ function (\ie, an $E_{k_j+1}^p$ function with
the bulk-to-boundary propagator of highest dimension changed by a bulk
propagator) and $p\leq k_j-2$. 
\begin{figure}[ht]
\begin{center}
\epsfxsize=12cm
\epsfbox{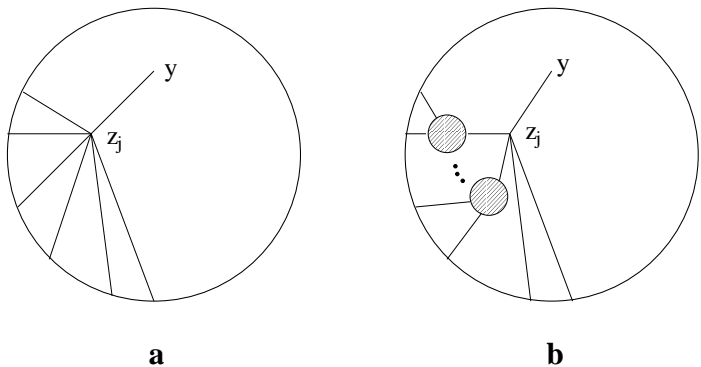}
\end{center}
\caption{Contact ($a$) and exchange ($b$) contributions to a
subdiagram of the exchange diagram in Fig.~10. \label{fig_subdiag}}
\end{figure}
This subdiagram can be either contact
or exchange (diagrams $a$ and $b$ of
Fig.~\ref{fig_subdiag}). The power counting shows that 
the integral over $z_j$ in the contact diagram is convergent.
This is all we need for diagram $a$. In diagram $b$
we can distinguish again two possibilites:
\begin{enumerate}
\item If the vertex at
$z_j$ is extremal with $\delta_j$ the highest dimension, according to
the analysis above one can only have a logarithmic divergence in the
integration over $z_j$ when $y\rightarrow x$ and $z_j \rightarrow
x$. The behaviour of the integrand is then 
\[
{y_0}^{(\sum_{l=1}^r
\delta_l + \sum_{i=1}^k \Delta_i -\Delta-1 -\eps^\prime)} \,
{(z_j)_0}^{\eps^\prime-1}, \]
where $\eps^\prime$ is a regulator
of the $z_j$ integral. The $1/\eps^\prime$ pole in the $z_j$ integral
is cancelled by a $\eps^\prime$ factor in the corresponding extremal
coupling. Moreover, as the vertex at $y$ is sub-extremal the
region $y\sim x$ of the integral over $y$ gives a vanishing
contribution. Therefore, the region $y\sim x$, $z\sim x$ does not
contribute to diagram $b$. On the other hand if $y$ is away
from $x$ there is 
no divergence in the $z_j$ integral, and the zero in the coupling
ensures that the diagram vanishes. 
\item If the vertex at $z_j$ is sub-extremal, the integral over $z_j$
converges. The divergences in the remaining integrals are cancelled by
zeros in the corresponding couplings. Therefore the diagram vanishes
if the coupling of the vertex at $z_j$ vanishes. As we show in the
text, if this coupling does not vanish the subdiagram contains at least one
(non-trivial) subdiagram that is a an $\tilde{E}_{k^\prime_j+1}^m$
function with $m\leq k^\prime_j - 2$ and the same procedure can be
employed to analyze it. 
\end{enumerate}
The iteration ends when the last $\tilde{E}$
function is a contact diagram.

\end{document}